\documentclass[fleqn,usenatbib]{mnras}

\usepackage[T1]{fontenc}

\DeclareRobustCommand{\VAN}[3]{#2}
\let\VANthebibliography\thebibliography
\def\thebibliography{\DeclareRobustCommand{\VAN}[3]{##3}\VANthebibliography}

\usepackage{graphicx}	
\usepackage{amsmath}	
\usepackage{amssymb}	
\usepackage{siunitx}
\usepackage{newtxtext,newtxmath}
\title[The Nature of High OIII/CII Galaxies]{The Nature of High [OIII]$_{\rm 88\mu m}$/[CII]$_{\rm 158\mu m}$ Galaxies in the Epoch of Reionization: Low Carbon Abundance and a Top-Heavy IMF?}

\author[H. Katz] {Harley Katz$^{1}$\thanks{E-mail:
  \href{mailto:harley.katz@physics.ox.ac.uk}{harley.katz@physics.ox.ac.uk}},
  Joakim Rosdahl$^{2}$,
  Taysun Kimm$^{3}$,
  Thibault Garel$^{4}$,
  J\'er\'emy Blaizot$^{2}$,
  \newauthor 
  Martin G. Haehnelt$^{5}$,
  L\'eo Michel-Dansac$^{2}$,
  Sergio Martin-Alvarez$^{5}$,
  Julien Devriendt$^{1}$,
  \newauthor
  Adrianne Slyz$^{1}$,
  Romain Teyssier$^{6}$,
  Pierre Ocvirk$^{7}$,
  Nicolas Laporte$^{5}$,
  and Richard Ellis$^{8}$\\ \\
  $^1$Sub-department of Astrophysics, University of Oxford,
  Keble Road, Oxford OX1 3RH, UK \\
  $^2$Univ Lyon, Univ Lyon1, Ens de Lyon, CNRS, Centre de Recherche
  Astrophysique de Lyon UMR5574, F-69230, Saint-Genis-Laval, France \\
  $^3$Department of Astronomy, Yonsei University, 50 Yonsei-ro,
  Seodaemun-gu, Seoul 03722, Republic of Korea \\  
  $^4$Observatoire de Geneve, Universite de Geneve, 51 Ch. des Maillettes, 1290 Versoix, Switzerland\\
  $^5$Kavli Institute for Cosmology and Institute of Astronomy,
  Madingley Road, Cambridge CB3 0HA, UK \\ 
  $^{6}$Institute for Computational Science, University of Zurich, Winterthurerstrasse 190, CH-8057 Zurich, Switzerland \\
  $^{7}$Observatoire Astronomique de Strasbourg, Université de Strasbourg, CNRS UMR 7550, 11 rue de l’Université, 67000 Strasbourg, France \\
  $^{8}$Department of Physics and Astronomy, University College London, Gower Street, London WC1E 6BT, UK
  }

\date{Accepted XXX. Received YYY; in original form ZZZ}

\pubyear{2021}

\begin{document}
\label{firstpage}
\pagerange{\pageref{firstpage}--\pageref{lastpage}}
\maketitle

\begin{abstract}
ALMA observations of $z>6$ galaxies have revealed abnormally high [OIII]$_{\rm 88\mu m}$/[CII]$_{\rm 158\mu m}$ ratios and [CII]$_{\rm 158\mu m}$ deficits compared to local galaxies. The origin of this behaviour is unknown. Numerous solutions have been proposed including differences in C and O abundance ratios, observational bias, and differences in ISM properties, including ionisation parameter, gas density, or photodissociation region (PDR) covering fraction. In order to elucidate the underlying physics that drives this high-redshift phenomenon, we employ {\small SPHINX$^{20}$}, a state-of-the-art, cosmological radiation-hydrodynamics simulation, that resolves detailed ISM properties of thousands of galaxies in the epoch of reionization which has been post-processed with {\small CLOUDY} to predict emission lines. We find that the observed $z>6$ [OIII]$_{\rm 88\mu m}$-SFR and [CII]$_{\rm 158\mu m}$-SFR relations can only be reproduced when the C/O abundance ratio is $\sim8\times$ lower than Solar and the total metal production is $\sim4\times$ higher than that of a Kroupa IMF. This implies that high-redshift galaxies are potentially primarily enriched by low-metallicity core-collapse supernovae with a more top-heavy IMF. As AGB stars and type-Ia supernova begin to contribute to the galaxy metallicity, both the [CII]$_{\rm 158\mu m}$-SFR and [CII]$_{\rm 158\mu m}$ luminosity functions are predicted to converge to observed values at $z\sim4.5$. While we demonstrate that ionisation parameter, LyC escape fraction, ISM gas density, and CMB attenuation all drive galaxies towards higher [OIII]$_{\rm 88\mu m}$/[CII]$_{\rm 158\mu m}$, observed values at $z>6$ can only be reproduced with substantially lower C/O abundances compared to Solar. The combination of [CII]$_{\rm 158\mu m}$ and [OIII]$_{\rm 88\mu m}$ can be used to predict the values of ionisation parameter, ISM gas density, and LyC escape fraction and we provide estimates of these quantities for nine observed $z>6$ galaxies. Finally, we demonstrate that [OI]$_{\rm 63\mu m}$ can be used as a replacement for [CII]$_{\rm 158\mu m}$ in high-redshift galaxies where [CII]$_{\rm 158\mu m}$ is unobserved and argue that more observation time should be used to target [OI]$_{\rm 63\mu m}$ at $z>6$. Future simulations will be needed to self-consistently address the numerous uncertainties surrounding a varying IMF at high redshift and the associated metal returns.
\end{abstract}




\section{Introduction}
Characterising the interstellar medium (ISM) of high-redshift galaxies deep into the epoch of reionization (EoR) represents one of the most interesting frontiers in the observational study of galaxy formation. Deep ALMA observations of forbidden infrared (IR) emission lines have now been confirmed out to $z>9$ \citep[e.g.][]{Hashimoto2018,Laporte2019}, providing insight into both the ISM properties and star formation histories of these primordial galaxies. As the sample of high-redshift systems with confirmed emission line detections increases, certain differences have been discovered between reionization-epoch galaxies and those in the local Universe.

The [CII]$_{\rm 158\mu m}$ line (hereafter [CII]), is one of the dominant coolants of the ISM and hence one of the brightest IR emission lines in the local Universe \citep[e.g.][]{Spitzer1978,Stacey1991}. With an ionisation potential of 11.3eV, [CII] emission can be stimulated in both ionised an neutral gas, often complicating its interpretation. Despite the wide range of ISM conditions that can emit [CII], it has been empirically shown to strongly correlate with star formation in local galaxies \citep[e.g.][]{Boselli2002,DeLooze2011,DeLooze2014,HerreraCamus2015}. Furthermore, observations have shown that the [CII] line is especially bright in low-mass metal-poor dwarf galaxies \citep[e.g.][]{Poglitsch1995,Cormier2010}, making it a particularly interesting target for studying the high-redshift Universe.

Earlier IR observations of $z>6$ galaxies focused primarily on this line due to its predicted high luminosity and observability with ALMA. There are now more than 20 galaxies at $z>6$ with [CII] detections as well as many more with non-detections. At slightly lower redshift (e.g. $4\lesssim z\lesssim6$), even larger samples of more than 100 galaxies have been surveyed \citep{LeFevre2020,Bethermin2020}, of which 75 have been detected in [CII] emission \citep{Schaerer2020}. The non-detections are of particular interest because this places many of the EoR galaxies far below the local relation between [CII] and the star formation rate (SFR) \citep[e.g.][]{Ota2014,Maiolino2015,Schaerer2015,Pentericci2016,Knudsen2016,Bradac2017,Laporte2019}. The exact origin of these deficits is currently unknown although some have proposed that the deficits are due to bursts of star formation \citep[e.g.][]{Ferrara2019}. However recent observations of extended, low surface brightness [CII] emission in the early Universe \citep{Fujimoto2019} has helped relieve some of the tension between the high-redshift galaxies and the low-redshift relations \citep{Carniani2020} in terms of absolute luminosity, but not surface brightness.

The [OIII]$_{\rm 88\mu m}$ line (hereafter [OIII]) is another strong IR emission line that has also been empirically established to correlate with star formation \citep[e.g.][]{DeLooze2014}. With an ionisation potential of 35.1eV, [OIII] only traces ionised gas, in contrast to [CII]. Interestingly, in metal-poor local Universe dwarf galaxies, [OIII] often outshines [CII] \citep[e.g.][]{Cormier2010,Madden2013}, representing up to 2\% of the total IR luminosity of the galaxy \citep{Cormier2015}. Since low-redshift metal poor dwarf galaxies are often considered analogues of high-redshift systems, [OIII] emission has been a particularly interesting target for the high-redshift Universe \citep[e.g.][]{Inoue2016} and has already been detected at $z>9$ \citep{Hashimoto2018}. 

While sample sizes of [OIII] emitting galaxies remain small at $z>6$, there have now been $\sim10$ detections, nearly all of which result in exceptionally high ratios of [OIII]/[CII], even compared to local metal-poor dwarf galaxies \citep[e.g.][]{Laporte2019,Harikane2020,Carniani2020}. Even when correcting for extended [CII] emission \citep{Carniani2020}, the [OIII]/[CII] ratios can reach as high as 8, which represents the upper tail of the distribution of low-redshift metal-poor galaxies \citep{Madden2013}. Similar to the [CII] deficits observed in high-redshift galaxies, the exact origin of the high [OIII]/[CII] ratios is still debated as the ionisation parameter, photodissociation region (PDR) covering fraction, enrichment patterns, metallicity, ISM density, and many other physical properties of a galaxy can impact the [OIII]/[CII] ratio \citep{Harikane2020}. 

Theoretical models, either analytic or numerical, are an invaluable tool for interpreting the origin of emission lines and physical properties of high-redshift galaxies. While historically, the understanding of emission lines, emission line ratios, and various strong-line diagnostics has relied upon idealised, 1D photoionization models \citep[e.g.][]{Kewley2002,Byler2017}, using codes such as {\small CLOUDY} \citep{Ferland2017} or {\small MAPPINGS} \citep{Sutherland2018}, these models often lack some of the complex physical processes (e.g. SN feedback, magnetic fields, shocks, density inhomogeneities, etc.) that affect the physics of the ISM. Although they are significantly more computationally expensive and it is difficult to run ``controlled" experiments as one can do with 1D photoionisation models, cosmological simulations provide the best insight into the complex physical nature of the high-redshift ISM. 

Due to their computational expense, most recent studies of IR emission lines in cosmological simulations have focused on small samples or individual galaxies \citep[e.g.][]{Vallini2015,Olsen2017,Katz2017,Katz2019,Pallottini2017,Pallottini2019,Arata2020,Lupi2019}, while only a few have attempted to model the emission in larger cosmological volumes \citep[e.g.][]{Moriwaki2018,Leung2020}. Obtaining larger samples of galaxies from simulations often relies on semi-analytic models \citep[e.g.][]{Popping2016,Lagache2018,Popping2019}. It is not clear whether the results from small sample sizes can be generalised to the larger galaxy population while semi-analytic models do not resolve much of the physics that dictates ISM properties. Furthermore, previous attempts at large volumes containing many galaxies often compromise on important physics such as radiation transfer or sacrifice resolution due to computational expense.

Not all results from simulations agree on the expected trends between [OIII] and [CII] emission and various galaxy properties, likely due to differences in spatial and mass resolution, included physics, sub-grid models, and the method for modelling emission lines. For example, the [OIII]-SFR relations found in \cite{Moriwaki2018,Arata2020} point towards larger [OIII] luminosities per unit SFR compared to \citet{Olsen2017,Katz2019}. Likewise, the few galaxies simulated in \cite{Katz2019} seem to fall closer to the local [CII]-SFR relation than those in \cite{Pallottini2017,Pallottini2019}. The origin of these discrepancies can in part be due to galaxy-to-galaxy variance, in addition to differences in modelling techniques, which are well established to be far from straightforward \citep{Olsen2018,Lupi2020}. Nevertheless, there have been many situations where simulations have agreed on emission-line properties of high-redshift galaxies and provide deep insights into or predictions of the early Universe. For example, the necessary but insufficient condition that LyC leakers have high O32 \citep{Katz2020,Barrow2020} as confirmed at $z\sim2-3$ \citep{Nakajima2020} or the strong [OI]$_{\rm 63\mu m}$ (hereafter [OI]) emission line \citep{Olsen2017,Katz2019}, that was recently observed at $z>6$ \citep{Rybak2020}.

While systematic uncertainties due to differences in modelling will continue to persist, issues due to galaxy-to-galaxy variance can be solved by simulating larger cosmological volumes. Until now there has been no full-box cosmological radiation hydrodynamics simulation with high enough resolution to semi-resolve the detailed structure of the ISM where these IR emission lines originate and with a large enough computational volume so that both the statistical properties of high-redshift galaxies can be studied and massive enough galaxies are probed to compare directly with observations. The primary goal of this work is to simulate the [CII] and [OIII] emission from such a state-of-the-art simulation to better understand the nature of high [OIII]/[CII] ratios at high redshift.

This paper is organised as follows: In Section~\ref{sims}, we briefly describe the {\small SPHINX$^{20}$} simulation (Rosdahl et al. {\it in prep.}) as well as our modelling technique for line emission. In Section~\ref{results}, we present the general [OIII] and [CII] properties of high-redshift galaxies at $z>4.68$ and discuss the physics that drives high [OIII]/[CII] ratios. Finally, in Section~\ref{conclusion}, we present our discussion and conclusions.

\section{Numerical Methods}
\label{sims}

\subsection{Cosmological Simulations}
This work makes use of the {\small SPHINX$^{20}$} cosmological radiation hydrodynamics simulation (Rosdahl et al. {\it in prep.}). {\small SPHINX$^{20}$} is the largest of all simulations in the {\small SPHINX} suite of cosmological radiation \citep{Rosdahl2018,Katz2020b} and magneto-radiation \citep{Katz2021} hydrodynamics simulations and is designed to model both the inhomogeneous process of reionization and the escape of LyC photons from the ISM. {\small SPHINX$^{20}$} is ideal for studying the emission line properties of high-redshift galaxies because for more than 20,000 galaxies at $z\geq4.68$, spanning a wide range in halo mass, stellar mass, and star-formation rate (SFR) in each simulation snapshot, we can probe the detailed properties of the ISM. This includes the inhomogeneous radiation field, as well as the detailed density structure, temperature, and metal enrichment, which represent key underlying quantities that impact emission line luminosities.

The details of {\small SPHINX$^{20}$} are very similar to those presented in \cite{Rosdahl2018}. We briefly describe the simulations with emphasis on the differences between the simulations used in this work compared to that of \cite{Rosdahl2018}. The simulations are run with {\small RAMSES-RT} \citep{Rosdahl2013,Rosdahl2015}, which is a radiation hydrodynamics version of the open-source, adaptive-mesh refinement {\small RAMSES} code \citep{Teyssier2002}. Initial conditions are generated with {\small MUSIC} \citep{Hahn2011} in a cosmological box with a comoving side length of 20~Mpc assuming the following cosmology: $\Omega_{\Lambda}=0.6825$, $\Omega_{\rm m}=0.3175$, $\Omega_{\rm b}=0.049$, $h=0.6711$, and $\sigma_8=0.83$. Due to the small volume of the simulations, the initial conditions were selected from multiple realisations to produce an average DM halo mass function at $z=6$ (see \citealt{Rosdahl2018}). The initial composition of the gas is set to be 76\% H and 24\% He by mass and have an initial metallicity of $3.2\times10^{-4}Z_{\odot}$. The dark matter particle mass is $2.5\times10^5{\rm M_{\odot}}$ and we have decreased the stellar particle mass from $1000{\rm M_{\odot}}$ in {\small SPHINX$^{10}$} to $400{\rm M_{\odot}}$ in {\small SPHINX$^{20}$}. The motivation for this change is to increase the stellar mass resolution to better capture star formation in lower mass galaxies. We can confirm that the stellar content of high mass galaxies is not significantly changed. The AMR grid is adaptively refined to resolve the local Jeans length by at least 4 cells or if the gas or DM mass of a cell is more than 8 times its initial value, resulting in a maximum physical resolution of $7.3{\rm pc}/h$ at $z=6$.

Stars are allowed to form in the simulation at the maximum level of refinement following a thermo-turbulent prescription \citep{Kimm2017,Trebitsch2017,Rosdahl2018}. Stellar feedback is included in the form of supernova (SN) explosions following a Kroupa stellar IMF \citep{Kroupa2001} and Lyman-continuum radiation. Star particles with $3{\rm Myr} < {\rm age} < 50 {\rm Myr}$ can explode via individual SNe by sampling from a delay-time distribution. When a SN event occurs, momentum is injected into the host and surrounding cells such that the amount of momentum injected is dependent on the phase of the SN that is resolved \citep{Kimm2015}. Star particles also continuously inject ionising photons into their host cells depending on their mass, age, and metallicity. In contrast to {\small SPHINX$^{10}$}, we only include two radiation energy bins to reduce the memory requirements of the simulation, set at the ionisation potentials of HI and HeI. The spectral energy distribution (SED) model employed in the simulation is {\small BPASS} \citep{Eldridge2008,Stanway2016}. We have updated the SED from {\small SPHINX$^{10}$} which uses v2.0 to v2.2.1 \citep{Stanway2018} and adopt a stellar IMF with a maximum mass of $100{\rm M_{\odot}}$, a minimum mass of $0.1{\rm M_{\odot}}$, that exhibits a broken power-law shape with a high mass slope of $-2.35$, a low mass slope of $-1.3$, and a break-mass of $0.5{\rm M_{\odot}}$ (very similar to a Kroupa IMF). Non-equilibrium chemistry is evolved locally for H, H$^+$, e$^-$, He, He$^+$, and He$^{++}$ as are the associated cooling processes (see the Appendix of \citealt{Rosdahl2013}). Cooling from metal lines is included at $T>10^4$K following tables computed with {\small CLOUDY} \citep{Ferland1996} as well as at $T\leq10^4$K using the analytic model from \cite{Rosen1995}.

For all simulation snapshots, we compute halo properties using the {\small ADAPTAHOP} halo finder \citep{Aubert2004,Tweed2009} in the most massive submaxima mode as described in \cite{Rosdahl2018}. Furthermore, LyC escape fractions are also computed for each halo following \cite{Rosdahl2018}, using the open-source {\small RASCAS} code \citep{Rascas2020}.

\subsection{Emission-Line Modelling}
Because our goal is to measure the [CII] and [OIII] emission line properties of high-redshift galaxies, we must compute the [CII] and [OIII] emission lines for each cell in the simulation. It is currently computationally intractable with our available resources to follow the full non-equilibrium chemistry and radiation coupling for C and O directly in the simulation so we resort to post-processing the simulations in order to measure the [CII] and [OIII] luminosities. We select seven simulation snapshots at $z=10,\ 9,\ 8,\ 7,\ 6,\ 5, \&\ 4.68$, the final redshift of our simulation for post-processing. Our method is similar in spirit to that used in \cite{Katz2019,Katz2019b,Katz2020}.

\begin{figure*}
\centerline{\includegraphics[scale=1,trim={0 0cm 0 0cm},clip]{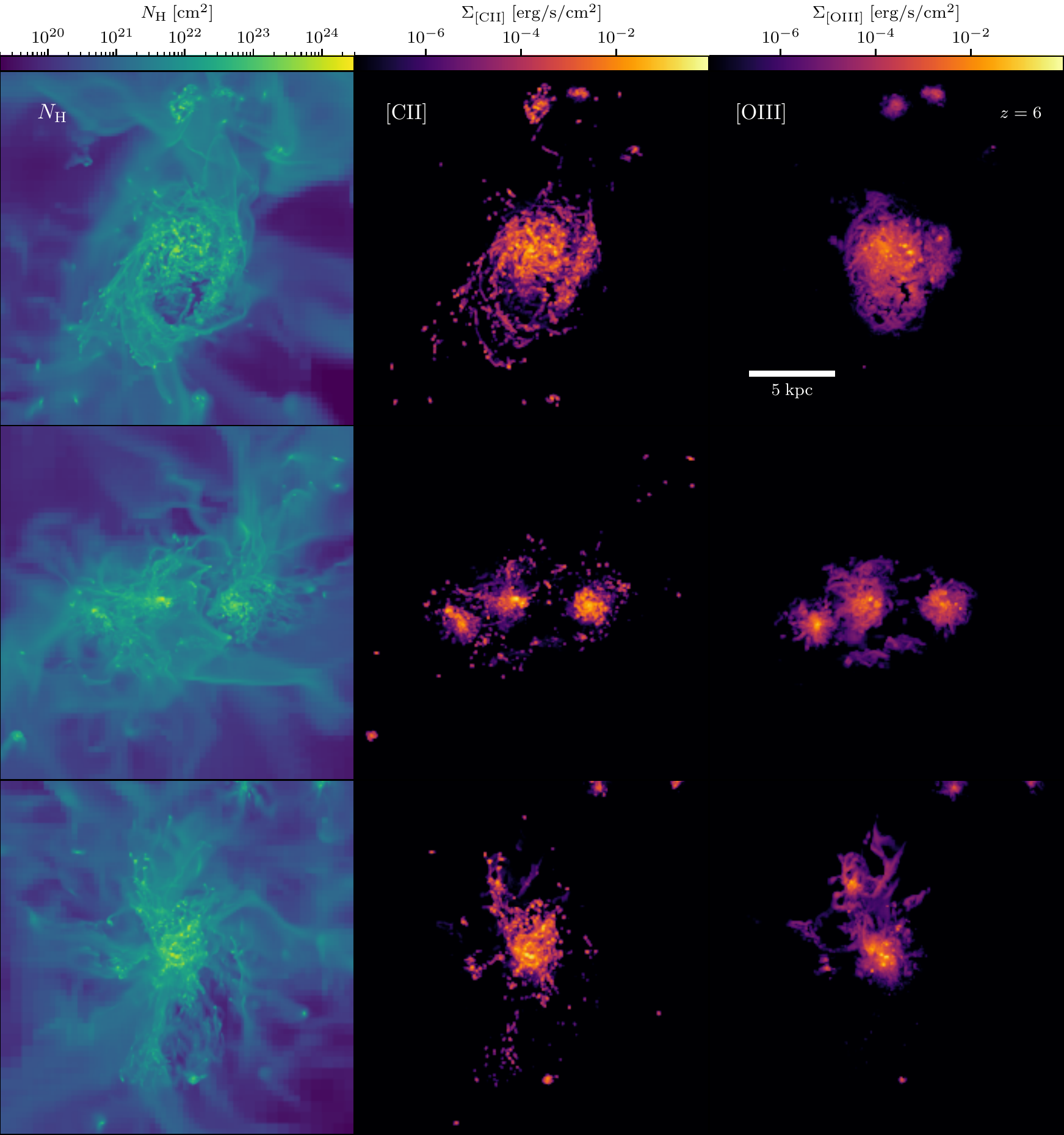}}
\caption{Hydrogen column density ($N_{\rm H}$), (left), [CII] surface brightness (centre), and [OIII] surface brightness (right) maps of the three most massive galaxies in {\small SPHINX$^{20}$}. These haloes have masses of $1.31\times10^{11}{\rm M_{\odot}}$, $1.29\times10^{11}{\rm M_{\odot}}$, and $9.97\times10^{10}{\rm M_{\odot}}$, and 10Myr averaged star formation rates (SFRs) of 17.6${\rm M_{\odot}yr^{-1}}$, 20.7${\rm M_{\odot}yr^{-1}}$, and 39.7${\rm M_{\odot}yr^{-1}}$, respectively. }
\label{hero}
\end{figure*}

For each cell in the simulation, we know the density, temperature, metallicity and ionisation parameter (the ratio of ionising photons to hydrogen atoms), as well as the cell size and redshift. For computational efficiency, the on-the-fly radiative transfer in the {\small SPHINX} simulations only follows H and He ionising photons (i.e. those with energies $>13.6$eV). To obtain the radiation field at energies below 13.6eV, we have post-processed the simulation snapshots with radiation transfer in the optically thin limit. This is done by restarting the simulation assuming that the gas, dark matter, and stars are static and using {\small RAMSES-RT} to evolve the radiation in two additional energy bins ($5.6{\rm eV}$-$11.2{\rm eV}$ and $11.2{\rm eV}$-$13.6{\rm eV}$) for at least the light-crossing time\footnote{Note that because the simulations use the variable speed of light approximation, we use a conservative estimate for the light crossing time by using the lowest value of the speed of light used in the simulation.} of the largest halo in the volume at that redshift. We then use {\small CLOUDY} \citep{Ferland2017} to compute the [OIII] and [CII] emission lines from the cells based on their density, temperature, metallicity, redshift, and radiation field in the four\footnote{This includes the two bins computed in post-processing and the two ionising radiation bins computed on-the-fly in the simulation.} energy bins, i.e. 8 local parameters.

Because there are billions of cells across the seven snapshots, it is impractical to compute a {\small CLOUDY} model for each cell. Furthermore, computing a grid across our eight parameters would mean that some parameters are only sparsely sampled in the grid which could result in errors in interpolation. Hence we adopt a different approach as follows:
\begin{enumerate}
    \item For each snapshot we randomly sample between 1.5-4.5 million simulation cells from galaxies across the full halo mass range.
    \item A {\small CLOUDY} model is run for the selected cells assuming an open geometry gas slab that has the same density, temperature, and metallicity as the simulation cell. The width of the slab is set to be the cell length and the shape of the radiation field within each radiation bin is set to follow that of the Milky Way (i.e. the total energy within the wavelength range of the radiation bin matches that of the simulation but the shape within the bin is kept constant for all cells). The {\small CLOUDY} models are all run in a constant temperature mode and a CMB background at the relevant redshift is also included.
    \item After the photoionisation models have been computed, for each snapshot, we randomly split the results into a training, cross-validation, and test set such that 25\% of cells are reserved for testing, 22.5\% are used for cross-validation, and 52.5\% of cells are used for training. For each line, we train two machine learning models. The first is a classifier that is used to predict whether the emission from the cell is $>10^{-3}L_{\odot}$ while the latter is used to predict the log of the total luminosity of the cell. In contrast to \cite{Katz2019}, we predict the log of the luminosity rather than luminosity to better the accuracy in the low luminosity regions which is important for extended emission. For this work, we use the {\small CATBOOST}\footnote{\url{https://github.com/catboost/catboost}} \citep{catboost2017,catboost2018} implementation of gradient boosted decision trees, which we find to be an improvement in terms of accuracy and well as training and prediction speed compared to the Random Forests used in \cite{Katz2019}. Models are trained for a maximum of 10,000 iterations and are stopped early if the loss-function does not improve for a period of 500 consecutive epochs. For the classifier, we use a binary cross-entropy loss function while for the regressor, we aim to maximise the $R^2$ score. For [CII], we find on the test set that the mean accuracy of the classifier is $99.75\pm0.08\%$, the mean $R^2$ score of the regressor is $0.996\pm0.001$, and that the total luminosity is conserved to within $2.89\%\pm2.84\%$. For [OIII], we find on the test set that the mean accuracy of the classifier is $99.91\pm0.06\%$, the mean $R^2$ score of the regressor is $0.997\pm0.002$, and that the total luminosity is conserved to within $11.7\%\pm6.53\%$. Our values of total luminosity conversion are well within the systematic errors of modelling the emission lines.
    \item Finally, we apply the models to every cell in each halo across all seven snapshots. 
\end{enumerate}

To run the {\small CLOUDY} models, we must also set the abundances of each metal. For our fiducial model, we use the {\small GASS} command that sets the abundance radios to be consistent with those of the Sun \citep{Grevesse2010}. Such abundances may not be representative of the early Universe due to different chemical enrichment scenarios \citep[e.g.][]{Maiolino2019}. For example, high-redshift galaxies may be predominantly enriched by core-collapse SN compared to a mix of core-collapse SN, type-Ia SN, and winds from AGB stars. For this reason, we also run an additional set of models where the chemical abundance ratios are consistent with the low metallicity (i.e. $1/20Z_{\odot}$)\footnote{Note that this value is higher than most of the galaxy metallicities in the simulation.} SN yields of \cite{Nomoto2006}\footnote{The stellar IMF used in \cite{Nomoto2006} has an upper mass threshold of $50{\rm M_{\odot}}$ rather than $100{\rm M_{\odot}}$ and a slope of $-2.35$ across the entire mass range of $0.07{\rm M_{\odot}}-50{\rm M_{\odot}}$, dissimilar to the IMFs used for stellar feedback in the simulation. Because of this difference, we have opted to keep the total metallicity fixed between the models using Solar and \cite{Nomoto2006} abundances patterns and only change the relative proportion of individual metals.}. This is particularly important for the [OIII]/[CII] ratio because the C/O abundance ratio from \cite{Nomoto2006} is eight times less than that of Solar abundance patterns. Note that the SED for all {\small CLOUDY} models is kept fixed for all three metal abundance models. This is further discussed in Section~\ref{caveats}. Finally, we also test models including molecules but find that our results are not substantially different when these are included and so for the results presented in this work, molecules have been excluded.

Our emission line modelling is computed for every halo in each snapshot with a halo mass $>10^8{\rm M_{\odot}}$. There are an average of 20,265 haloes of this mass across the seven snapshots.

\begin{figure}
\centerline{\includegraphics[scale=1,trim={0 0.4cm 0 1.5cm},clip]{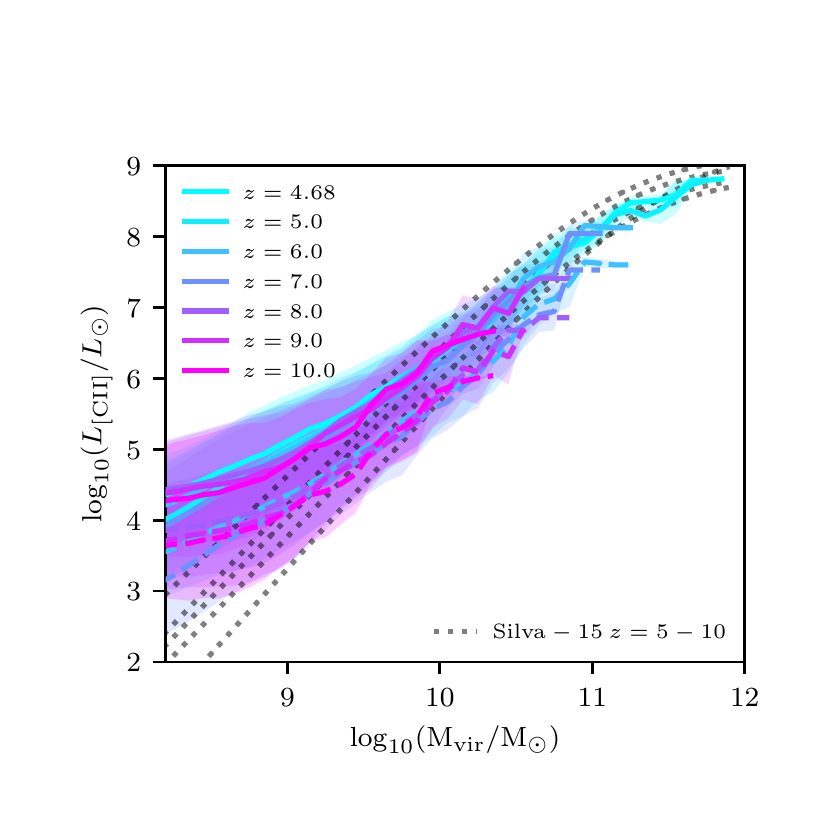}}
\centerline{\includegraphics[scale=1,trim={0 0.4cm 0 1.5cm},clip]{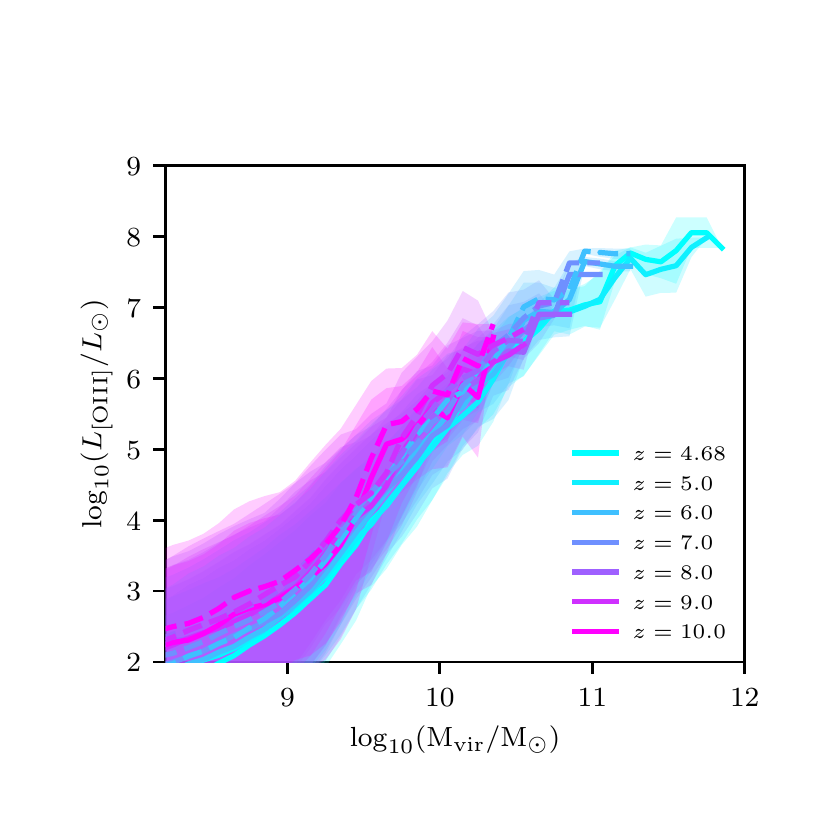}}
\caption{[CII]-${\rm M_{vir}}$ (top) and [OIII]-${\rm M_{vir}}$ (bottom) relations for galaxies in {\small SPHINX$^{20}$} at $z=4.68-10$. The solid and dashed lines represent the results for the Solar abundance and core collapse SN abundance models, respectively. The shaded regions represent the $1\sigma$ scatter about the relation. The dotted lines in the top panel represent the semi-analytic models from \protect\cite{Silva2015}.  Little evolution is seen in these relations as a function of redshift whereas systematic differences occur due to choice of abundance model.}
\label{lum_mvir}
\end{figure}

\begin{figure}
\centerline{\includegraphics[scale=1,trim={0 0.4cm 0 1.5cm},clip]{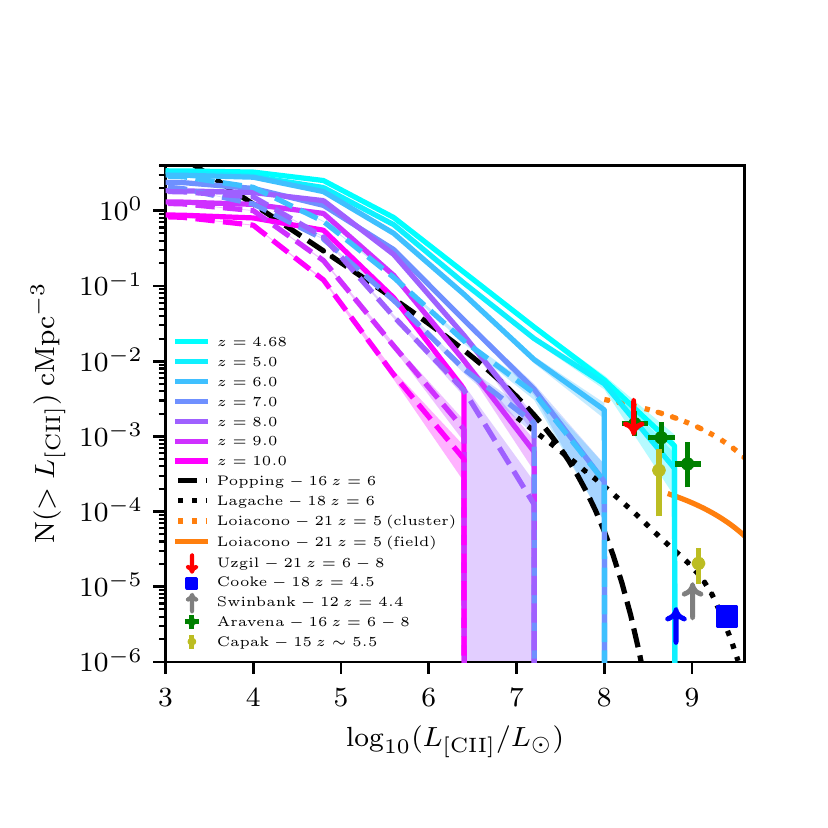}}
\centerline{\includegraphics[scale=1,trim={0 0.4cm 0 1.5cm},clip]{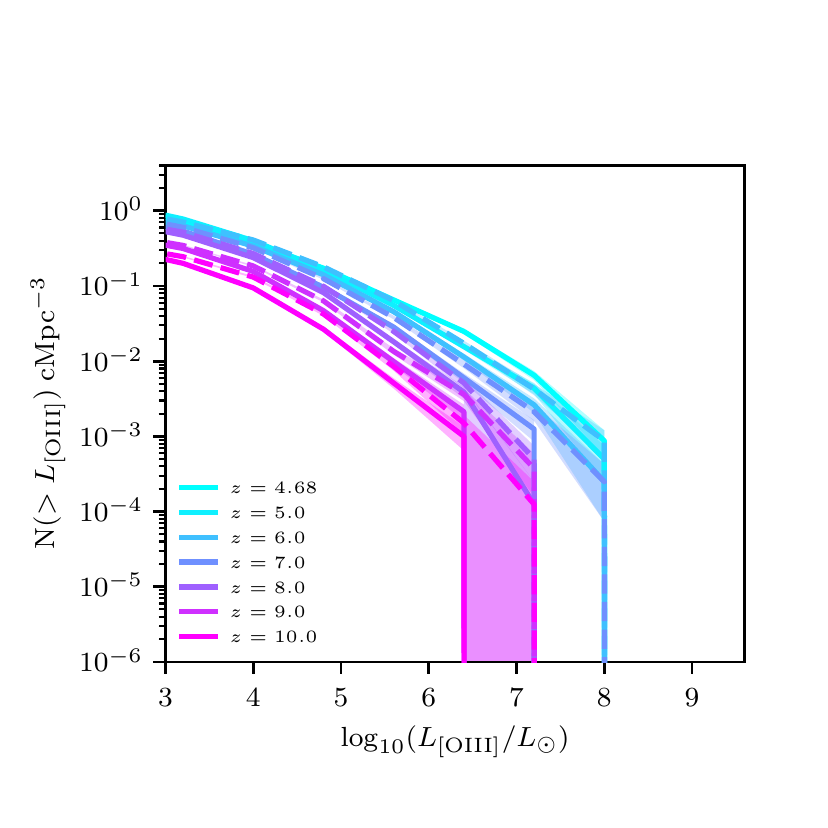}}
\caption{[CII] (top) and [OIII] (bottom) luminosity functions at different redshifts for galaxies in {\small SPHINX$^{20}$} versus observations. The solid and dashed lines represent the results for Solar abundance and core collapse SN abundance models, respectively. The shaded region represents the $1\sigma$ scatter about the relation. Observational constraints for the [CII] luminosity function are shown for \protect\cite{Swinbank2012,Capak2015,Aravena2016,Cooke2018,Uzgil2021,Loiacono2021} and the dotted and dashed lines represent results from the semi-analytic models of \protect\cite{Lagache2018} and \protect\cite{Popping2016}, respectively. No observational constraints are available for the [OIII] luminosity function at high redshift.}
\label{lum_funcs}
\end{figure}

\section{Results}
\label{results}
In this section, we present our results on [CII] and [OIII] emission from high-redshift galaxies, with a particular focus on the origin of high [OIII]/[CII] ratios at $z>6$ and how these two emission lines can be used to constrain the properties of high-redshift galaxies. 

We have calculated the [CII] and [OIII] emission for thousands of galaxies in each simulation snapshot between $z=10$ and $z=4.68$. In Figure~\ref{hero} we show density, [CII], and [OIII] maps of the three most massive haloes at $z=6$, with halo masses of $1.31\times10^{11}{\rm M_{\odot}}$, $1.29\times10^{11}{\rm M_{\odot}}$, and $9.97\times10^{10}{\rm M_{\odot}}$, respectively. These galaxies have 10Myr averaged star formation rates (SFRs) of 17.6${\rm M_{\odot}yr^{-1}}$, 20.7${\rm M_{\odot}yr^{-1}}$, and 39.7${\rm M_{\odot}yr^{-1}}$, well within the range of galaxies at have been observed at $z>6$ \citep[e.g.][]{Maiolino2015}.

These galaxies exhibit very different morphologies. The first is a turbulent, and partially disrupted rotating disc, the second is a triple major merger, and the third, while more similar to the first, possesses two strongly star forming regions and clear evidence of dynamical disturbance. Such properties are rather characteristic of high-redshift galaxies where strong dynamical interactions are more common than in lower-redshift galaxies.

Consistent with our earlier work \citep{Katz2019}, we find that the [CII] emission follows the densest gas structures in the galaxy while [OIII] exhibits a different morphology, more coincident with the young star forming regions and mildly lower density gas (see also \citealt{Pallottini2019a,Lupi2020}). With a higher ionisation potential and lower critical density compared to C$^+$, such behaviour is expected for O$^{++}$.

\subsection{General Properties}
With such a large sample of galaxies, we can begin to overcome issues related to sample variance and make predictions regarding the relations between [CII] and [OIII] luminosity and galaxy properties.

\begin{figure*}
\centerline{\includegraphics[scale=1,trim={0 0 0 0},clip]{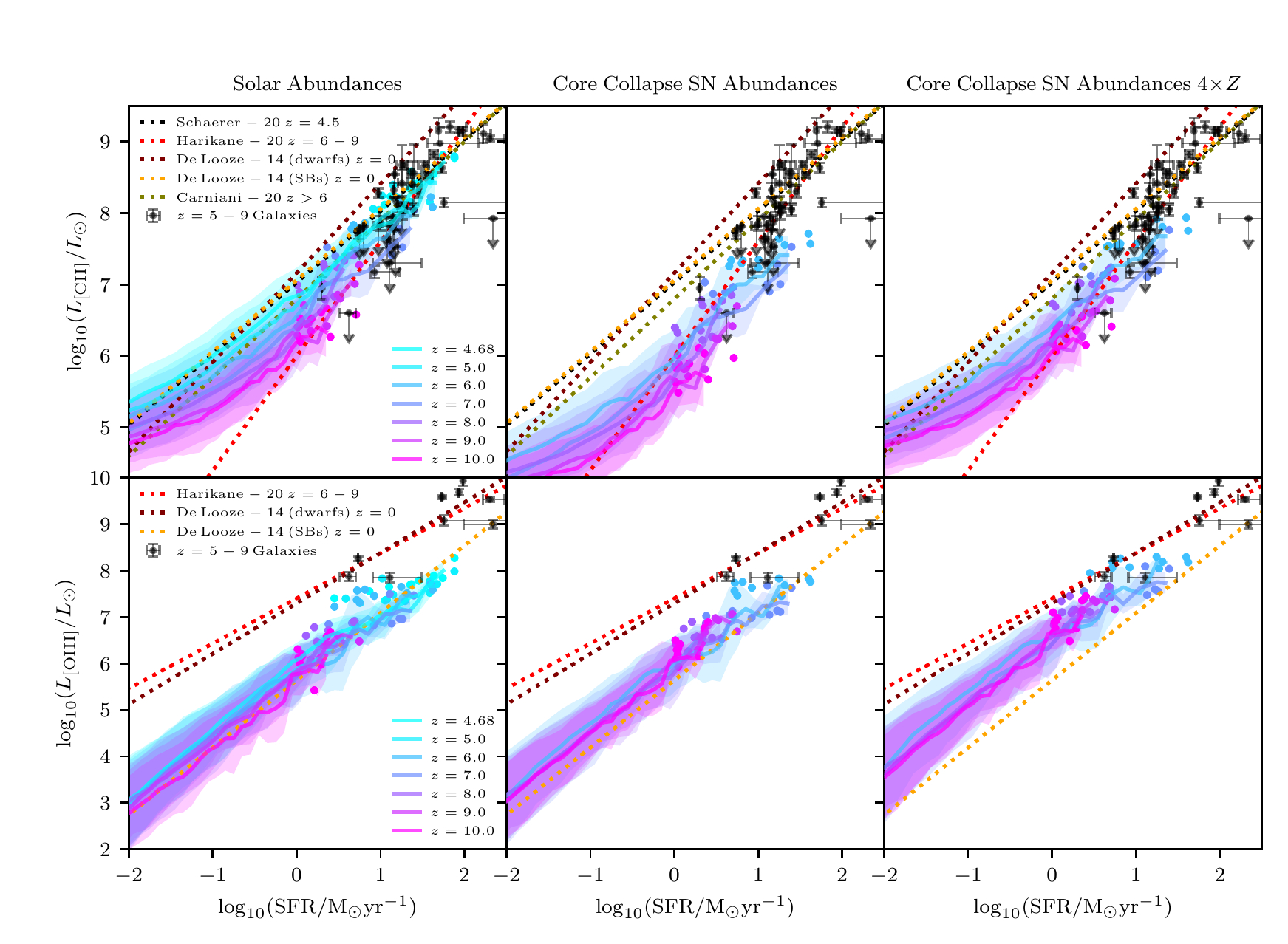}}
\caption{Mean [CII]-SFR relations (top) and [OIII]-SFR relations (bottom) for galaxies in the {\small SPHINX$^{20}$} simulation at $z=5-9$ compared to observations. The left, centre, and right rows represent models where we assume Solar abundances, core collapse SN abundances, or core collapse SN abundances with 4 times enhanced metallicity, respectively. The shaded regions represent the $1\sigma$ scatter about the relation. Black data points with errors represent observed galaxies at $z=5-9$ whereas coloured data points represent the brightest [CII] and [OIII] emitters at each redshift. We show high-redshift fits from \protect\cite{Schaerer2020,Harikane2020,Carniani2020} as well as $z=0$ relations for dwarfs and starbursts (SBs) from \protect\cite{DeLooze2014}.}
\label{CII-SFR}
\end{figure*}

\subsubsection{${\rm [CII]-M_{vir}}$ and ${\rm [OIII]-M_{vir}}$ relations}
In Figure~\ref{lum_mvir} we show the relation between [CII] and [OIII] luminosity and halo virial mass as a function of redshift for the models with Solar abundance ratios and core collapse SN abundance ratios. As halo mass increases, so do the [CII] and [OIII] luminosities. Our predicted relation between $L_{\rm [CII]}$ and ${\rm M_{vir}}$ using Solar abundance ratios is in good agreement with the semi-analytic model of \cite{Silva2015} at ${\rm M_{vir}\gtrsim9.5M_{\odot}}$. The two models deviate from each other at the low mass end. Our models with the same abundance ratios exhibit almost no redshift dependence. Similarly, the relation between $L_{\rm [OIII]}$ and ${\rm M_{vir}}$ shows no evidence of redshift dependence. 

Constraining the [CII]-${\rm M_{vir}}$ and [OIII]-${\rm M_{vir}}$ relations is of particular importance for forecasting intensity mapping results \citep[e.g.][]{Lagache2018,Yue2019} and we have tables of these values available online\footnote{\url{https://github.com/HarleyKatz/SPHINX-CII-OIII}}. 

\subsubsection{${\rm [CII]}$ and ${\rm [OIII]}$ luminosity functions}
Since ${\rm M_{vir}}$ is not an observed quantity, we do not have any direct observational constraints between [OIII], [CII], and ${\rm M_{vir}}$. However, observational constraints are available for the [CII] luminosity function and in the top panel of Figure~\ref{lum_funcs}, we compare our predicted [CII] luminosity functions as a function of redshift to various observational constraints \citep{Swinbank2012,Capak2015,Aravena2016,Cooke2018,Uzgil2021} as well as other semi-analytic models \citep{Popping2016,Lagache2018}. We observe a strong redshift dependence of the [CII] luminosity function such that as redshift decreases, the number of [CII] emitters at all luminosities increases. Our simulations have very few galaxies with [CII] luminosities $>10^{42}{\rm erg\ s^{-1}}$ which is approximately the lower limit of where observational constraints begin (e.g. at $z=4.68$ there are 5 such galaxies while there are none at $z=6$). Nevertheless, we find that the $z\sim5.5$ constraint from \cite{Capak2015} falls in between our $z=5$ and $z=6$ predictions. Our models predict far fewer [CII] emitters than the estimates from \cite{Aravena2016} at $z=6-8$, and are more in line with the $z=6-8$ upper limit from \cite{Uzgil2021}.  Compared to results from commonly used semi-analytic models, we predict significantly more [CII] emitters at $L_{\rm [CII]}>10^{8}L_{\odot}$ than both \cite{Popping2016,Lagache2018}. The model that uses core collapse SN abundance ratios predicts a luminosity function much lower than observations.

Unfortunately, there are no observational constraints on the [OIII] luminosity function at high redshift and it is not often discussed in the literature (although see \citealt{Moriwaki2018}). Once again, we expect less redshift evolution in the [OIII] luminosity function compared to [CII] due to the fact that O is predominantly produced in core collapse SN; hence, at very high redshifts (e.g. $z\gtrsim9$), [OIII] may be the most likely FIR line to be observed.

\subsubsection{${\rm [CII]-SFR}$ and ${\rm [OIII]-SFR}$ relations}
There has been a considerable amount of debate in the literature, both from an observational and theoretical perspective on how the high-redshift [CII]-SFR relation compares to that in the local Universe.  Early observations suggested a large deficit of [CII] emission in $z>6$ galaxies compared to those at $z=0$ \citep[e.g.][]{Maiolino2015,Pentericci2016,Laporte2019}. However, more recent observations that account for extended [CII] emission reduce some of this difference \citep{Carniani2020} and by $z\sim4.5$, the observed relation is nearly identical to that in the local Universe \citep{Schaerer2020}. 

The origin of [CII] deficits is currently unknown as simulations also do not agree. For example, the zoom-in simulations of individual high-redshift galaxies in \cite{Katz2019} show that haloes tend to fall very close to the observed low-redshift relation, and when [CII] deficits do occur in their simulations, it is after a burst of star formation, when the galaxy is also at its brightest, perhaps suggesting an observational bias. The simulations of \cite{Arata2020} produce a [CII]-SFR relation steeper than that observed in the local Universe, likely due to different enrichment channels as well as different ISM properties, although their highly star-forming galaxies are consistent with $z=0$ observations. In contrast, \cite{Pallottini2017,Pallottini2019} find that for similar halo masses to \cite{Katz2019}, their haloes have both higher star formation rates and significantly larger [CII] deficits.

This disagreement is also present between different simulations regarding the high-redshift [OIII]-SFR relation where \cite{Katz2019,Olsen2017} find [OIII] deficits while \cite{Arata2020,Moriwaki2018}, find an [OIII]-SFR relation consistent with $z=0$. Interestingly, while \cite{Arata2020} cannot produce lower luminosity [OIII] emitters, \cite{Katz2019} struggles to produce high [OIII] emitters. In summary, the current lack of agreement in the literature, both in observations and simulations, prohibits robust interpretations of observed high-redshift galaxies and deserves further study.

\begin{figure}
\centerline{\includegraphics[scale=1,trim={0 0.4cm 0 1.5cm},clip]{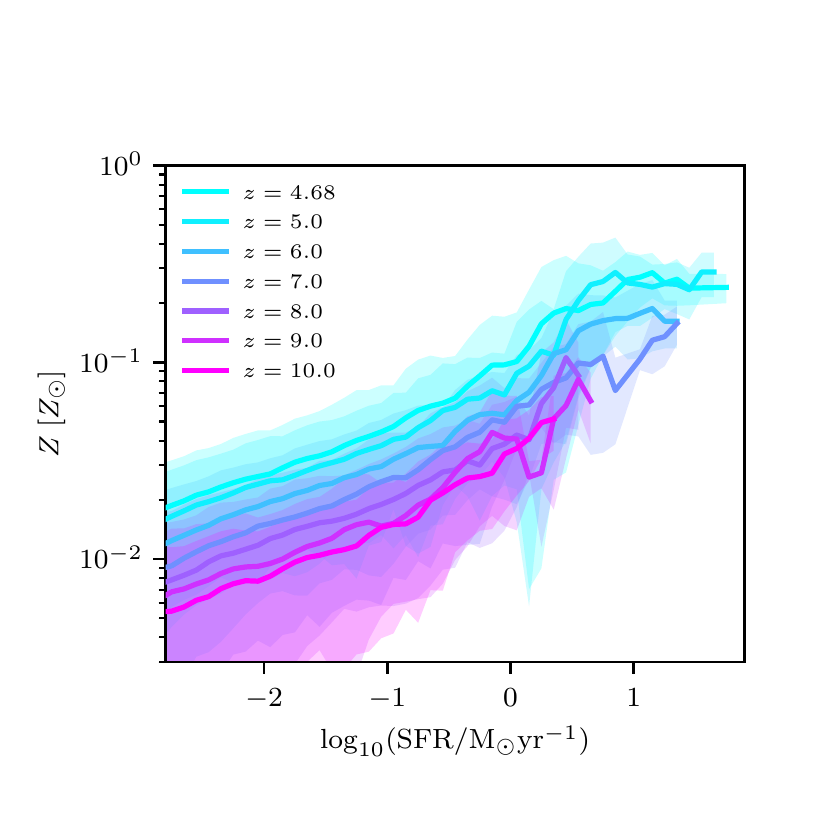}}
\caption{SFR-$Z$ relation as a function of redshift. $Z$ represents the mass-weighted gas-phase metallicity. The shaded region represents the $1\sigma$ standard deviation about the relation. As redshift decreases, galaxies at fixed SFR have higher metallicities.}
\label{sfr_zism}
\end{figure}

\begin{figure}
\centerline{\includegraphics[scale=1,trim={0 0.4cm 0 1.5cm},clip]{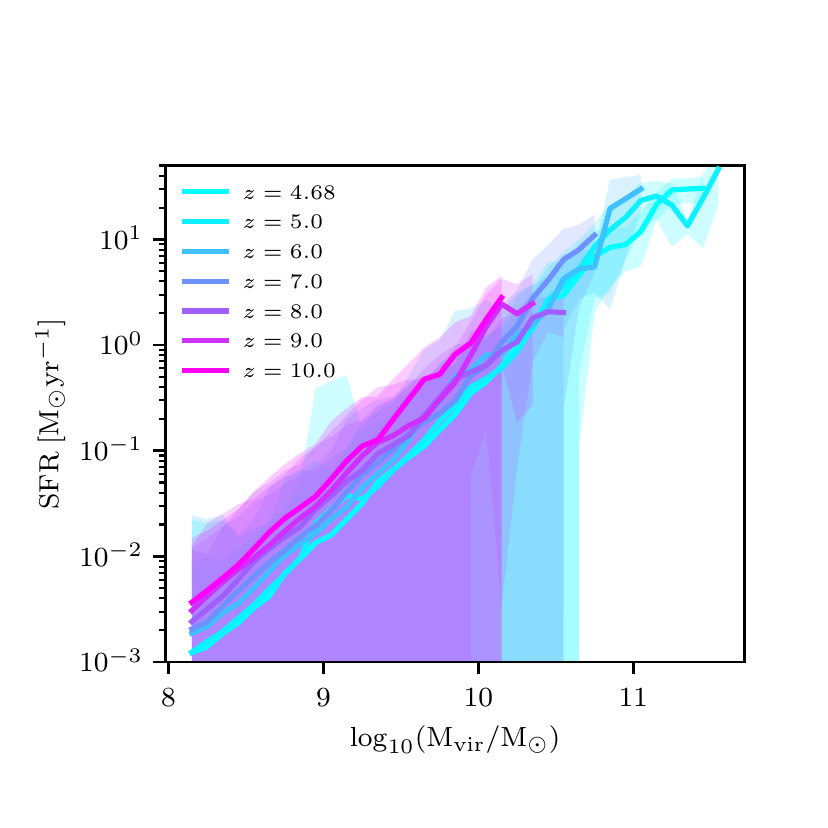}}
\caption{SFR-${\rm M_{vir}}$ relation as a function of redshift. The shaded region represents the $1\sigma$ standard deviation about the relation. As redshift decreases, galaxies at fixed SFR have higher halo masses or alternatively, galaxies at fixed halo mass have lower SFR.}
\label{sfr_mvir}
\end{figure}

In the top row of Figure~\ref{CII-SFR}, we show the [CII]-SFR relations for {\small SPHINX$^{20}$} galaxies at different redshifts compared with observations. We show models using Solar abundances, core collapse SN abundances, and core collapse SN abundances where we have artificially and uniformly enhanced the metallicities of the gas cells in each simulated galaxy by a factor of 4. Regardless of the abundance model, we see a strong evolution with redshift in that at fixed SFR, the [CII] luminosity increases with decreasing redshift. To first order, the line luminosity scales with the absolute abundance of the metal which is why the Solar abundance model produces higher [CII] for fixed SFR compared to the core collapse abundance model. In Figure~\ref{sfr_zism}, we plot the mean ISM metallicity as a function of SFR for various redshifts and see that as redshift decreases, the ISM metallicity at fixed SFR increases. Hence we would expect an increase in [CII] luminosity. The [CII] luminosity is also sensitive to the amount of low temperature neutral gas. We expect more of this gas is present in higher halo masses.  In Figure~\ref{sfr_mvir} we show the evolution of the ${\rm M_{vir}}$-SFR relation as a function of redshift and we also see here that at fixed SFR, halo mass increases with decreasing redshift. Hence this effect should also drive higher [CII] luminosities both due to the increased amount of gas and the decreased susceptibility of the galaxy to SN feedback.  

Since [OIII] is related to the efficiency of the feedback in the galaxy, we would expect the opposite effect compared to [CII]. The larger halo mass at fixed SFR would reduce the amount of [OIII] emission and counteract the increase in metallicity with decreasing redshift. The bottom row of Figure~\ref{CII-SFR} shows the [OIII]-SFR relations as a function of redshift compared with observations. Regardless of the choice of abundance ratios, there is almost no redshift evolution. While it is expected that this relation does not evolve as strongly with redshift as the [CII]-SFR relation, it is an interesting coincidence that there is no redshift evolution at all. 

The redshift evolution in the [CII]-SFR relation may be even more pronounced than shown in Figure~\ref{CII-SFR} due to the expected evolution in C abundance. While O is predominantly formed in core collapse SNe soon after the onset of star formation, the formation timescale for C is significantly longer due to the fact that it is expelled into galaxies via both core collapse SNe and type-Ia SNe as well as in winds from asymptotic giant branch (AGB) stars \citep[e.g.][]{Maiolino2019}. Hence the amount of C per unit metallicity is expected to increase with decreasing redshift and therefore enhance the redshift evolution in the relation. The top left and top centre panels of Figure~\ref{CII-SFR} show the extreme cases of Solar abundances and core collapse SN abundances, respectively. Since we do not explicitly model the varying timescales of enrichment of different elements in the simulation, we have modulated the abundance ratios in post-processing to explore how this physics may affect the [CII]-SFR relation. Interestingly, when we assume Solar abundance ratios, we find that the simulated galaxies are in much better agreement with observations than when we assume core collapse SN abundance ratios. At $z\geq6$, our Solar abundance relations are very consistent with the fitted relation from \cite{Harikane2020}, and by $z=4.68$, our relation has converged to that of \cite{Schaerer2015}. At ${\rm SFR<1M_{\odot}yr^{-1}}$, our relations seem to converge more towards those of \cite{Carniani2020} than that of \cite{Harikane2020}, although at higher SFR, these two relations are relatively indistinguishable.

Compared to the local relation (shown as the yellow and brown lines in Figure~\ref{CII-SFR}), our $z\geq6$ galaxies tend to produce small [CII] deficits even for Solar abundance ratios possibly due to the lower metallicity and lower halo mass per fixed SFR. The most [CII] bright galaxies in {\small SPHINX$^{20}$} are however consistent with the local relation when assuming Solar abundance ratios. This is consistent with the results found in \cite{Katz2019} which used similar subgrid physics and spatial/mass resolution and demonstrates the need for larger samples of galaxies.

When assuming core collapse SN abundance ratios, we find very strong [CII] deficits that appear inconsistent with the high-redshift galaxy population due to the drastically reduced C abundance. Taken at face-value, it would appear that our simulations prefer galaxies to exhibit abundance ratios closer to Solar values. However, we argue below that this is not necessarily the case. 

One can see in the bottom row of Figure~\ref{CII-SFR} that regardless whether we assume Solar abundances or core collapse SN abundances, our [OIII]-SFR relation underpredicts individual high-redshift observations as well as the fitted relation from \cite{Harikane2020}. There is little difference between the [OIII]-SFR relation for the two abundance assumptions because the primary O enrichment channel is via core collapse SN and thus the O abundance is very similar between the two models. Hence while a Solar abundance ratios model can reproduce the [CII]-SFR relation at high redshift, it struggles with the [OIII]-SFR relation, as was the case in our earlier work \cite{Katz2019}. How then can we reconcile both relations with observations?

Changing the ISM properties of our high redshift galaxies is unlikely to relieve the tension. Because [OIII] and [CII] originate in different regions of temperature-density phase-space, attempting to create more [OIII]-emitting gas would reduce the amount of [CII]-emitting gas. Hence for the Solar abundance model, the reduction in [CII] would disrupt the current agreement in favour of improving the [OIII]-SFR relation. For this reason, this is not a preferred solution. If the galaxies were significantly gas richer at fixed halo mass in reality compared to what is predicted by the simulation, then this solution could be an option. However, \cite{Katz2020} showed that for the {\small SPHINX} suite of simulations, these high-mass galaxies accrete gas relatively efficiently. The gas mass could only increase by a factor of about two as to not supersede the cosmic baryon fraction, which is not nearly enough to overcome the near order of magnitude deficit in the [OIII]-SFR relation.

\begin{figure*}
\centerline{\includegraphics[scale=0.98,trim={0 0cm 0 0cm},clip]{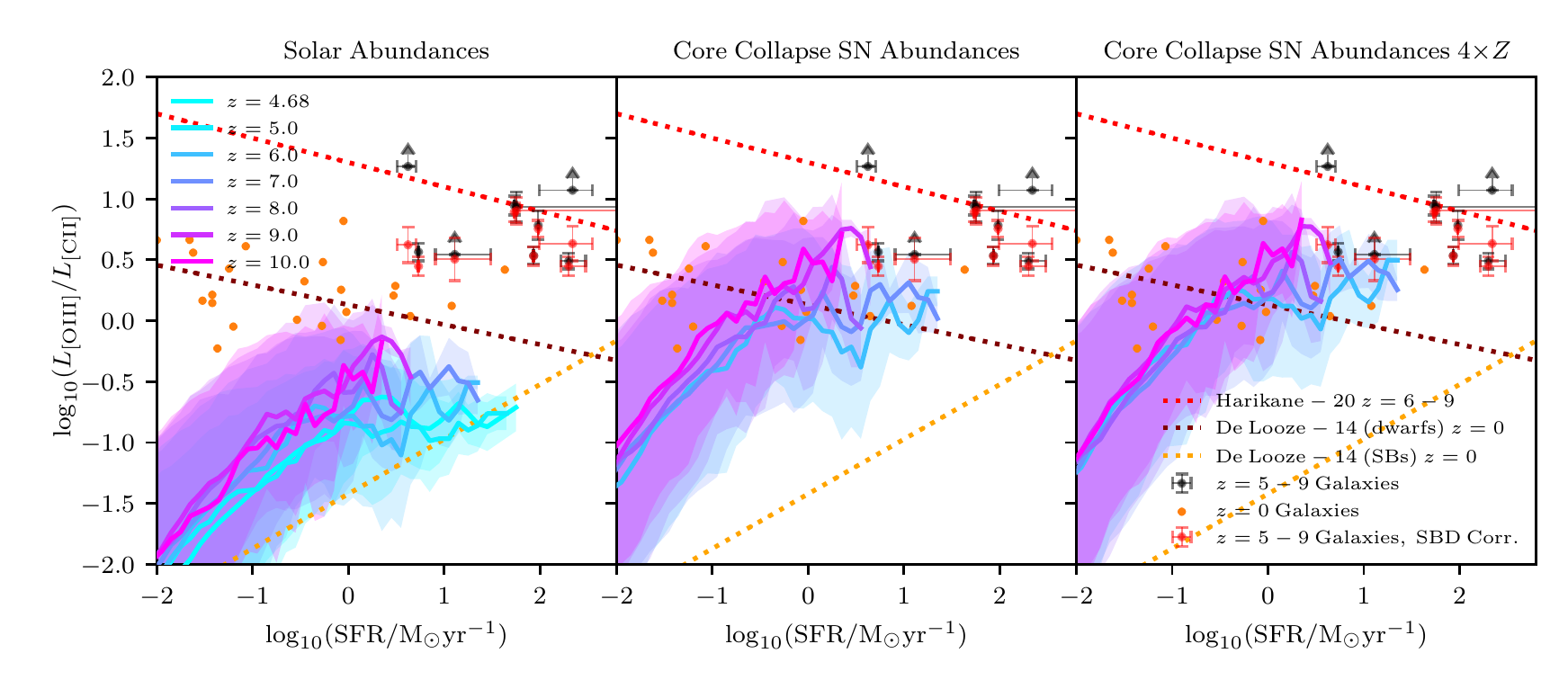}}
\caption{[OIII]/[CII] versus SFR for galaxies in our simulation at different redshifts compared to observations for the Solar abundance model (top) and the core collapse SN abundance model (bottom). The shaded regions represent the $1\sigma$ scatter about the relation. Black data points represent $z=5-9$ galaxies from \protect\cite{Harikane2020} while red data points represent those same galaxies accounting for surface brightness dimming \protect\citep{Carniani2020}. Orange points represent individual low redshift dwarf galaxies from the DGS survey \protect\citep{Madden2013}. The red dotted line is the high-redshift fitted relation from \protect\cite{Harikane2020} while the maroon and orange lines are the $z=0$ relations for dwarf and starburst galaxies, respectively, from \protect\cite{DeLooze2014}.}
\label{CII-OIII-SFR}
\end{figure*}

\begin{figure*}
\centerline{\includegraphics[scale=1,trim={0 0 0 0},clip]{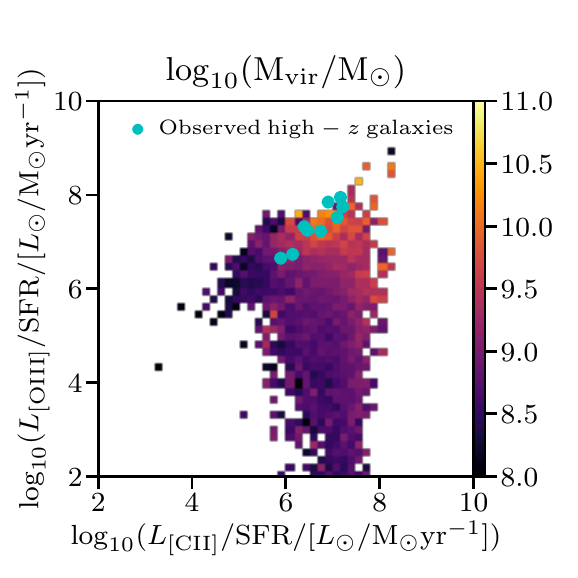}\includegraphics[scale=1,trim={0 0 0 0},clip]{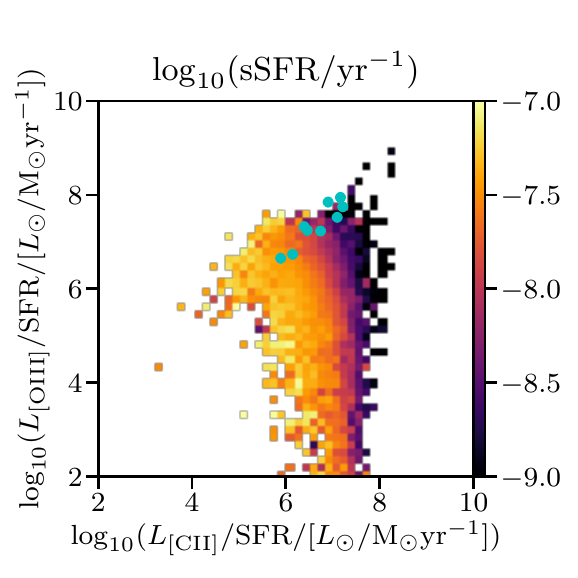}\includegraphics[scale=1,trim={0 0 0 0},clip]{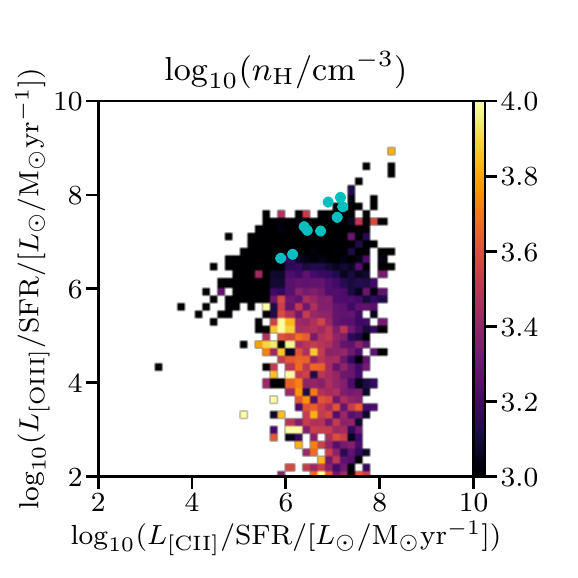}}
\centerline{\includegraphics[scale=1,trim={0 0 0 0},clip]{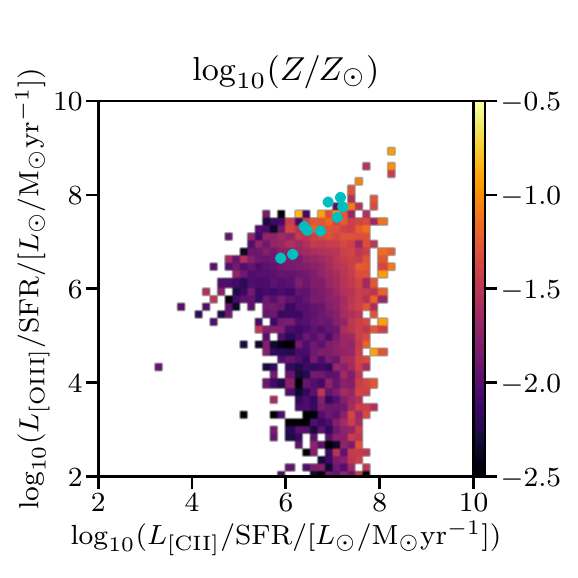}\includegraphics[scale=1,trim={0 0 0 0},clip]{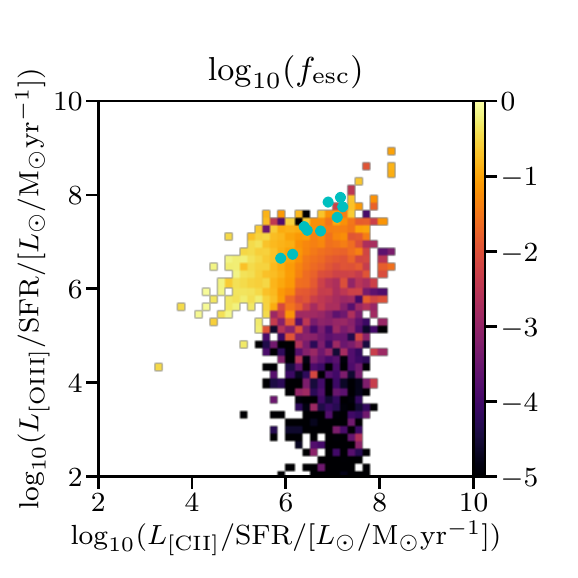}\includegraphics[scale=1,trim={0 0 0 0},clip]{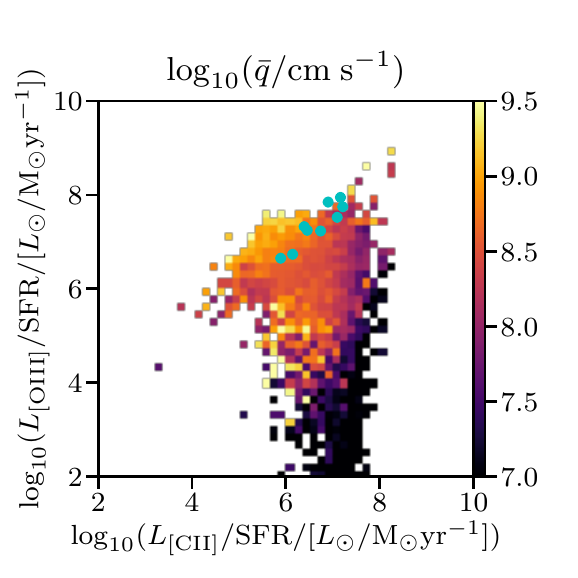}}
\caption{2D histograms of [CII]/SFR versus [OIII]/SFR for galaxies in the {\small SPHINX$^{20}$} simulation between $6\leq z\leq10$ coloured by virial mass (top left), sSFR (top centre), 90th percentile ISM density (top right), ISM metallicity (bottom left), LyC escape fraction (bottom centre), and 90th percentile ISM ionisation parameter (bottom right). We show results for the core collapse SN abundance model with the 4$\times$ enhanced metallicity. $n_{\rm H}$, $Z$, and $q$ have are calculated as the mass-weighted quantities in the [CII] and [OIII] emitting regions ($Z$, $n_{\rm H}$) or HII regions ($\bar{q}$). Cyan points represent individual high-redshift galaxies from \protect\cite{Carniani2020}.}
\label{2Dhists}
\end{figure*}

Since when assuming Solar abundance ratios, the [CII]-SFR relation is in good agreement with observations, we argue that it is important that the total carbon mass in galaxies is not significantly different from our fiducial model. One solution to our problem could be to increase the abundances of all elements for the \cite{Nomoto2006} abundance patterns. To explore this option, in the right panel of Figure~\ref{CII-SFR}, we plot the relations where we have re-run CLOUDY models where the total metallicity is increased by a factor of 4. Here we can see that the good agreement between the [CII]-SFR relation is once again recovered while the [OIII]-SFR relation is in significantly better agreement with the observed high-redshift galaxies. We emphasise that in this model, the C/O relation is the same as in the core collapse SN abundance model, but the total amount of metals is increased. 

While this solution may seem {\it ad hoc}, there is precedent for this in high-redshift galaxies. Simulations show that low metallicity galaxies may produce a more top-heavy stellar initial mass function (IMF) compared to what is observed in the local Universe \citep[e.g.][]{Safranek2014,Chon2021}. More top-heavy IMFs result in an increased amount of SNe which leads to an increase in metal production as well as more high-energy photons which can form more O$^{++}$. \cite{Chon2021} show that the high-mass stellar IMF slope can become significantly flatter at stellar metallicities in the range $0.10Z_{\odot}<Z<0.1Z_{\odot}$. While the most massive galaxies in our simulation have gas-phase metallicities slightly less than $0.2Z_{\odot}$ at $z=6$, we highlight the fact that only the nebular metallicity must be enhanced for our proposed solution to work, not the stellar metallicity. \cite{Steidel2016} showed that in order to explain the SEDs of certain $z\sim2-3$ galaxies, the nebular metallicities appear $5\times$ enhanced compared to stellar metallicities. This is partly driven by the way in which stellar and nebular abundances are measured, for example using different elements (e.g. O and Fe) to measure each. Any true enhancement in the nebular metallicity compared to stellar metallicity will support our argument. Furthermore, the work of \cite{Chon2021} and \cite{Safranek2014} assume solar abundance patterns. At low temperatures, their cooling functions are dictated by [CII] emission. If one assumes that metal abundances are more consistent with \cite{Nomoto2006}, low temperature cooling is likely suppressed which could lead to a top-heavy IMF at even higher metallicities than reported in their work. We emphasise again that given the uncertainties in both the modelling in this work and in determining the stellar IMF from first principles in simulations, both our factor of 4 boost and the discussion in this section are intended to be qualitative rather than quantitative. In {\small SPHINX$^{20}$}, we already artificially boost the energy from SNe by a factor of 4 in order to match the high-redshift UV luminosity function \citep{Rosdahl2018}. This extra energy is thought to be needed in order to overcome numerical over-cooling; however it also may be reflecting the necessity for a more top-heavy IMF. Although the energy from SN in the simulation is artificially boosted, the metal yields are still consistent with a standard \citep{Kroupa2001} IMF. A more top-heavy IMF may produce an ISM that is similar to what we find in this work already due to the boosted feedback although this neglects the fact that the additional metals may enhance cooling. Therefore, we argue that our results are potentially indicative of a top-heavy IMF in the early Universe. 

\cite{Arata2020} have explicitly shown with simulations that as metallicity increases, the O/C ratio drops due to enrichment from AGB stars and type Ia SNe. Thus the high-redshift metal-poor galaxies in their work naturally exhibit high [OIII]/[CII] ratios. In contrast to our results where our fiducial model has difficulty producing bright [OIII] emitters, their model struggles in predicting lower luminosity [OIII] and [CII] emitters at high redshift. However since they only simulate a few haloes, the origin of this discrepancy is inconclusive. \cite{Arata2020} choose a slightly more top-heavy IMF (Chabrier) than used in the fiducial model of this work (Kroupa). Their halo at $z=6$ is more massive than any of the haloes in {\small SPHINX} and thus a deeper potential well may allow the object to retain more of its metals, although their feedback is more efficient at regulating the stellar mass of the object. Perhaps the largest difference between the two simulations is that \cite{Arata2020} does not include cooling below $10^4$K which is important for a multi-phase ISM and emergent line luminosities. This discrepancy makes it difficult to compare the line emission from the two simulations because the [CII] emission is highly sensitive to gas temperature.

\subsubsection{${\rm [OIII]/[CII]}$ ratios}
In Figure~\ref{CII-OIII-SFR} we plot mean [OIII]/[CII] versus SFR relations for {\small SPHINX$^{20}$} galaxies compared with observations. We show the Solar abundance model (left), the core collapse SN abundance model (centre), and the core collapse SN abundance model where the metallicity has been enhanced by a factor of 4 (right). Note that the latter two models predict very similar [OIII]/[CII] ratios. Thus any future discussion of [OIII]/[CII] ratios in the core collapse SN abundance model also applies to that with boosted metallicity.

In contrast to the observed relations, we find that the trend between [OIII]/[CII] and SFR is not monotonic. Between $-2<\log_{10}({\rm SFR/M_{\odot}yr^{-1}})<0$, we find decreasing [OIII]/[CII] ratios as SFR decreases however, at $\log_{10}({\rm SFR/M_{\odot}yr^{-1}})>0$, the trend flattens, regardless of the assumption on abundance ratios. Redshift evolution pushes galaxies towards lower [OIII]/[CII] with decreasing redshift, especially at $z<6$. 

As expected, compared to the observed high-redshift, $z=5-9$ galaxies, the Solar abundance model produces [OIII]/[CII] ratios that are far lower. At SFRs of $10{\rm M_{\odot}yr^{-1}}$ the [OIII]/[CII] deficit for the mean galaxy in our simulations is more than an order of magnitude compared to observations (see also \citealt{Laporte2019}). The [OIII]/[CII] ratios are even lower than the metal poor dwarf galaxies in the local Universe and more consistent with local starburst galaxies. However, when we assume that the O and C abundances are similar to that of core collapse SNe, we find that the high [OIII]/[CII] ratios at high redshift can be reproduced by our simulation. Still, we do not have many galaxies with ${\rm [OIII]/[CII]>10}$ as seen in \cite{Harikane2020}. When accounting for [CII] surface brightness dimming corrections \citep{Carniani2020}, we find very good agreement between $z=5-9$ galaxies and the highly star-forming galaxies in {\small SPHINX$^{20}$}. 

Our results indicate that the C/O abundance ratios are a crucial factor for high-redshift [OIII]/[CII] ratios. Such an argument was proposed by \cite{Steidel2016} to explain the offsets of $z\sim2-3$ galaxies on the BPT diagram. Similarly, \cite{Arata2020} demonstrated with cosmological simulations that explicitly track C and O abundances from different chemical enrichment channels that the C/O ratio is indeed significantly lower in high-redshift galaxies due to the delayed onset of AGB winds. However, \cite{Harikane2020} suggest that while lower C/O ratios may explain some of the enhanced [OIII]/[CII] ratios observed at high redshift, the reduced C abundance alone is not enough to explain some of the non-detections of [CII]; hence, the ratio by itself cannot explain the galaxies with $>1$~dex lower [CII]/SFR. Once again, these [CII] deficits are partially relieved by surface brightness dimming corrections \citep{Carniani2020}; however, it is important to understand what other physical mechanisms can drive high [OIII]/[CII] ratios as well as [CII] deficits.

\subsection{What else drives high [OIII]/[CII] ratios?}
\cite{Harikane2020} systematically discuss eight different effects that can impact the [OIII]/[CII] ratio in high-redshift galaxies. They employ idealised models of the ISM using {\small CLOUDY} to better understand the impact of different ISM properties on the [OIII]/[CII] ratio. While these simplified photoionization models are useful for isolating individual effects, real high-redshift galaxies exhibit a complicated, multi-phase ISM structure. In this section, we explore seven of the eight effects\footnote{The last effect is the lower C/O abundance ratios due to different enrichment channels in the early Universe which has already been discussed above.} in the context of the {\small SPHINX$^{20}$} simulations.

\subsubsection{Lower PDR covering fractions}
[OIII] is generally emitted very close to stars from HII regions while [CII] tends to originate further out in PDRs. Therefore if the HII region becomes large enough so that the gas cloud is density-bounded (i.e. the outer radius of the cloud is smaller than the Stromgren radius), we expect there to be very little [CII] emission. Molecular clouds exhibit complex density structures where some regions around the star are expected to remain shielded from the ionising radiation and exist in a neutral state. \cite{Harikane2020} argue that if the covering fraction of PDRs ($C_{\rm PDR}$) surrounding stars is significantly less than unity, we would expect an increase in [OIII]/[CII].  

Because the ISM structure in our simulations is highly complex, we cannot define a direct metric for $C_{\rm PDR}$ in our simulations that matches the idealised models of \cite{Harikane2020}. Nevertheless, we have measured the LyC escape fraction ($f_{\rm esc}$) for each galaxy in {\small SPHINX$^{20}$} which provides a proxy for $C_{\rm PDR}$ on a galaxy scale. We expect that galaxies that have higher $f_{\rm esc}$ have lower PDR covering fractions and should thus exhibit higher ratios of [OIII]/[CII].

From simulations, it is well established that $f_{\rm esc}$ is a feedback regulated quantity \citep[e.g.][]{Trebitsch2017,Kimm2017}. Hence it is the SN feedback that dictates the PDR covering fraction in our simulation more than the ionising photon production. Such an effect can only be realised in numerical simulations.  

In the bottom-centre panel of Figure~\ref{2Dhists}, we show a 2D histogram of [CII]/SFR versus [OIII]/SFR coloured by $\log_{10}(f_{\rm esc})$. Galaxies in our simulation that exhibit the highest values of $f_{\rm esc}$ and, by deduction, lower PDR covering fractions also tend to exhibit weaker [CII]/SFR and higher [OIII]/SFR, or in other words, higher [OIII]/[CII] ratios.

As the Universe evolves and becomes more metal enriched, the typical mass of a galaxy also increases while the luminosity-weighted global LyC escape fractions tend to decrease \citep[e.g.][]{Rosdahl2018}. Certain analytical models predict that $f_{\rm esc}$ evolves very strongly with redshift such that $f_{\rm esc}\propto(1+z)^{3.4}$ \citep{Haardt2012}. \cite{Harikane2020} show that for $C_{\rm PDR}=0$, [CII] luminosity is decreased by 99\% and therefore, the evolving escape fraction can contribute to the high [OIII]/[CII] ratios seen at high redshift. Furthermore, given the correlation between [OIII]/[CII] and $f_{\rm esc}$, \cite{Inoue2016,Katz2020} argued that one can select LyC leakers from non-leakers based on this ratio alone. This metric was tentatively tested \citep{Katz2020} on the few low-redshift galaxies where observations of both lines exist and a common trend for the systems that were predicted to be leakers was that many appeared to have strong signatures of feedback, consistent with the results from our simulations.

\begin{figure}
\centerline{\includegraphics[scale=1,trim={0 0.4cm 0 1.5cm},clip]{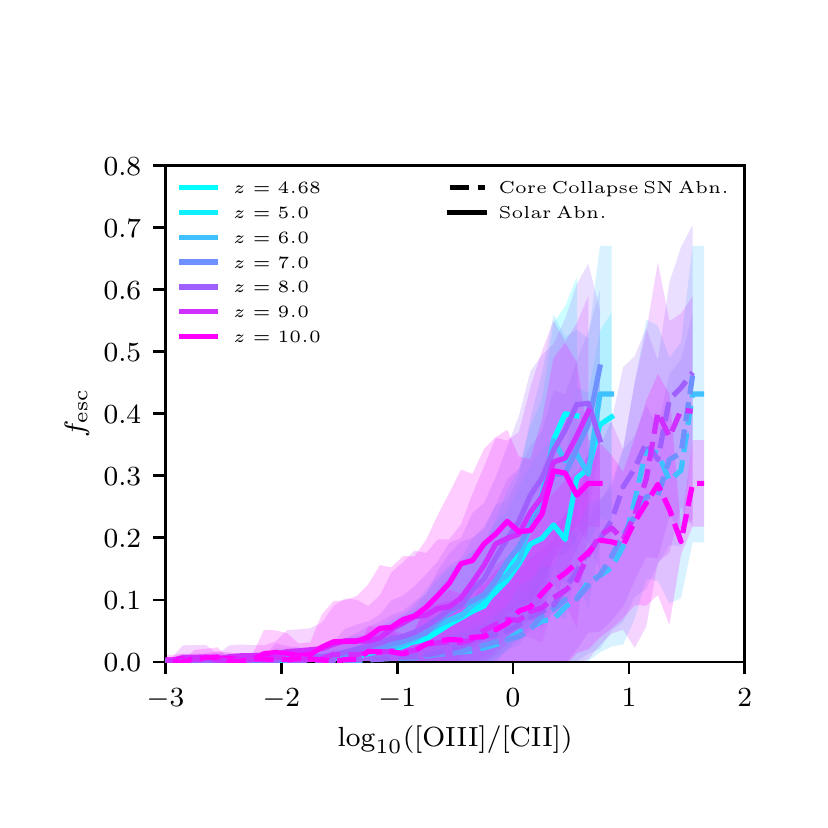}}
\caption{[OIII]/[CII] versus $f_{\rm esc}$ for galaxies in our simulation versus observations. The coloured lines and shaded regions indicate the mean relation and $1\sigma$ scatter for different redshift bins. Solid and dashed lines represent the results when we assume Solar abundances and core collapse SN abundances, respectively. We only include galaxies in this plot that have an ${\rm SFR}>10^{-2}{\rm M_{\odot}yr^{-1}}$.}
\label{CII-OIII-desc}
\end{figure}

To demonstrate this relation more explicitly, in Figure~\ref{CII-OIII-desc} we plot the mean value and $1\sigma$ scatter of $f_{\rm esc}$ versus $\log_{10}({\rm [OIII]/[CII]})$ for each redshift interval for models where we assume either Solar abundance ratios or core collapse SN abundance ratios. Although there is considerable scatter in the relation, there is a very strong trend that the galaxies that have the highest values of [OIII]/[CII] also exhibit the highest $f_{\rm esc}$. Assuming core collapse SN abundances systematically shifts the relation towards higher values of [OIII]/[CII]; nevertheless, the trend persists. 

Consistent with \cite{Harikane2020}, we find that lower PDR covering fractions or equivalently increased LyC escape fractions can drive galaxies towards higher [OIII]/[CII].

\subsubsection{Higher ISM densities}
Galaxies that exhibit the highest ISM densities are expected to exhibit the lowest ratios of [CII]/SFR and [OIII]/SFR because once the density surpasses the critical density of the line, the emissivity increases as $n$ rather than $n^2$ due to collisional de-excitation. In the top right panel of Figure~\ref{2Dhists} we show a 2D histogram of [CII]/SFR versus [OIII]/SFR, coloured by the 90th percentile density for gas that emits\footnote{This gas is defined as having $T<10^5$K and $\rho>1{\rm cm^{-3}}$.} [CII] and [OIII]. There is a clear gradient indicating that the galaxies with the lowest values of [CII]/SFR and [OIII]/SFR also have the highest ISM densities.

\begin{figure}
\centerline{\includegraphics[scale=1,trim={0 0.4cm 0 1.5cm},clip]{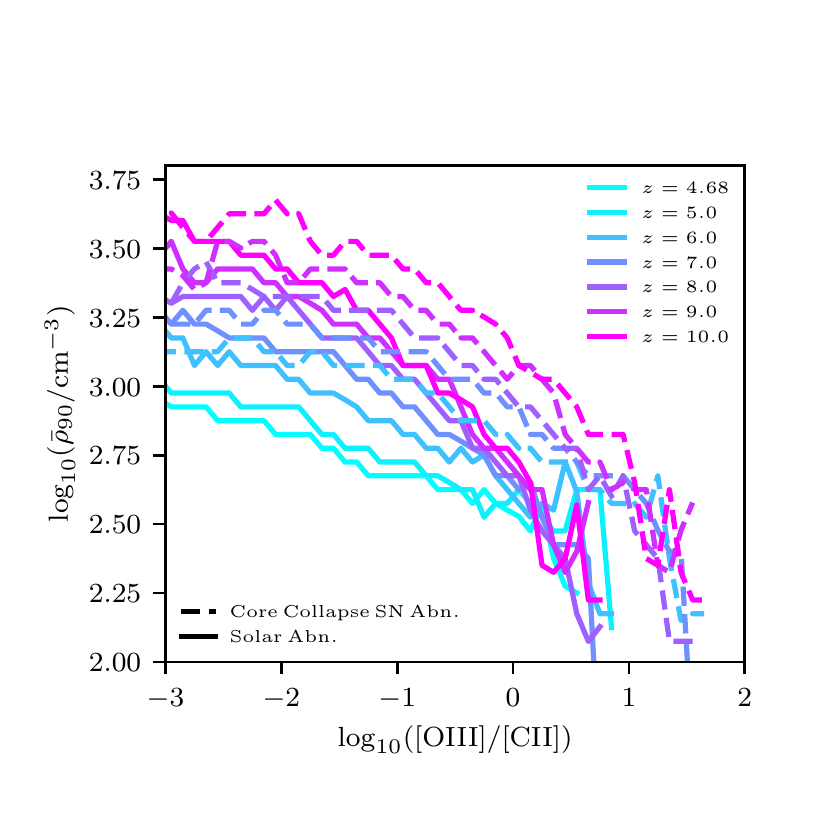}}
\caption{[OIII]/[CII] versus $\log_{10}(\bar{\rho}_{90}/{\rm cm^{-3}})$ for galaxies in our simulation versus observations. $\rho$ is defined to be the 90th percentile value from the [CII] and [OIII] emitting gas in the galaxies. The coloured lines indicate the median relation for different redshift bins. Solid and dashed lines represent the results when we assume Solar abundances and core collapse SN abundances, respectively. We only include galaxies in this plot that have an ${\rm SFR}>10^{-2}{\rm M_{\odot}yr^{-1}}$.}
\label{CII-OIII-rho}
\end{figure}

The critical density of [CII] for interactions with hydrogen is $\sim2800{\rm cm^{-3}}$ while the critical density of [OIII] for interactions with electrons is $\sim510{\rm cm^{-3}}$ \citep{Carilli2013}. Hence we expect that galaxies with higher ISM densities will exhibit lower ratios of [OIII]/[CII] because as density increases, we reach the critical density of [OIII] before [CII]. In Figure~\ref{CII-OIII-rho}, we plot  the 90th percentile ISM density of the [OIII] and [CII] emitting gas in the simulated galaxies versus [OIII]/[CII] for the cases where we assume both Solar abundances as well as core collapse SN abundances. For lower values of [OIII]/[CII], we see a decrease in ISM density with decreasing redshift. Consistent with expectations, galaxies with lower ISM densities exhibit higher ratios of [OIII]/[CII]. 

We do however stress that one should exercise caution when interpreting the trend between ISM density and [OIII]/[CII] for our simulated high-redshift galaxies. While such effects can be easily tested in controlled idealised models such as those presented in \cite{Harikane2020}, in cosmological simulations, other parameters are not held fixed with the changing ISM density. For example, low density may indicate that the galaxy has been susceptible to strong SN feedback which can destroy the [CII] emitting regions of the galaxy, enhance the amount of O$^{++}$, and increase $f_{\rm esc}$, all of which are important for setting the [OIII]/[CII] ratio. Furthermore, it is well established that [OIII] is emitted from different regions of temperature-density phase-space than [CII] and thus the gas that emits [OIII] may be characterised by a different density compared to that which emits [CII].

\begin{figure}
\centerline{\includegraphics[scale=1,trim={0 0.4cm 0 1.5cm},clip]{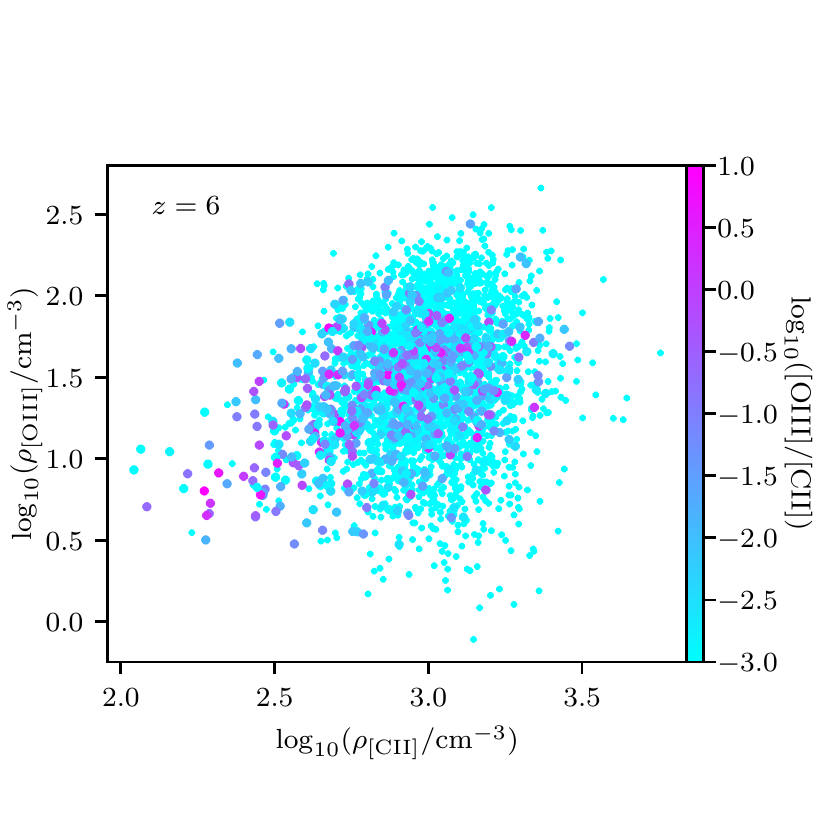}}
\caption{Luminosity-weighted gas density of [CII] regions versus [OIII] regions for galaxies at $z=6$ coloured by the ratio of [OIII]/[CII]. We only include galaxies that have ${\rm SFR}>10^{-2}{\rm M_{\odot}yr^{-1}}$ as well as [CII] and [OIII] luminosities $>10^3L_{\odot}$. [OIII]/[CII] ratios are shown for the Solar abundance model although the results are not fundamentally different for the core collapse SN abundance model.}
\label{rho_cii_oii}
\end{figure}

To demonstrate this, in Figure~\ref{rho_cii_oii} we plot the luminosity-weighted mean gas density for [CII] against that of [OIII] coloured by the ratio of [OIII]/[CII]. While there is considerable scatter, in general, the galaxy regions that emit strongly in [OIII] have significantly lower gas densities compared to the regions with high [CII] luminosity. It is often the case that the densities are more than an order of magnitude lower in the [OIII] emitting regions compared to [CII]. This suggests that idealised 1D photoionization models with a fixed ISM density poorly mimic the conditions of [OIII] and [CII] emitting regions in the ISM of high-redshift galaxies. Rather, our simulations are more akin\footnote{This statement may appear to conflict with the {\small CLOUDY} models that were used to post-process the cosmological simulations in this work which assumed constant density and temperature. However, the ISM in our simulations is semi-resolved so each individual model is more similar to an individual zone in a {\small CLOUDY} model that captures HII regions, and PDRs in the same slab.} to a constant pressure assumption for the 1D models (see e.g. \citealt{Cormier2019}) that can, in principle, capture these density contrasts. Since, many of the galaxies with high [OIII]/[CII] ratios do not have ISM densities above the critical density for [OIII] and [CII], despite the large density contrasts between [CII] and [OIII] emitting regions, the higher ISM densities do not play a strong role in setting the [OIII]/[CII] ratio.

\subsubsection{Higher ionisation parameters}
The photoionization models of \cite{Harikane2020} show that models with higher ionisation parameters ($q=cn_{\gamma,{\rm ion}}/n_{\rm H}$) tend to exhibit higher ratios of [OIII]/[CII]. In our simulations, [CII] is predominantly emitted from higher density neutral regions whereas [OIII] originates in lower density, ionised gas with $T\gtrsim10^4$K. Additional high-energy photons can ionise some of the C$^+$ into C$^{++}$ which also serves to reduce [CII] emission while additional hard ionising photons have the potential to create more O$^{++}$ thereby increasing the ratio of [OIII]/[CII]. 

\begin{figure}
\centerline{\includegraphics[scale=1,trim={0 0.4cm 0 1.5cm},clip]{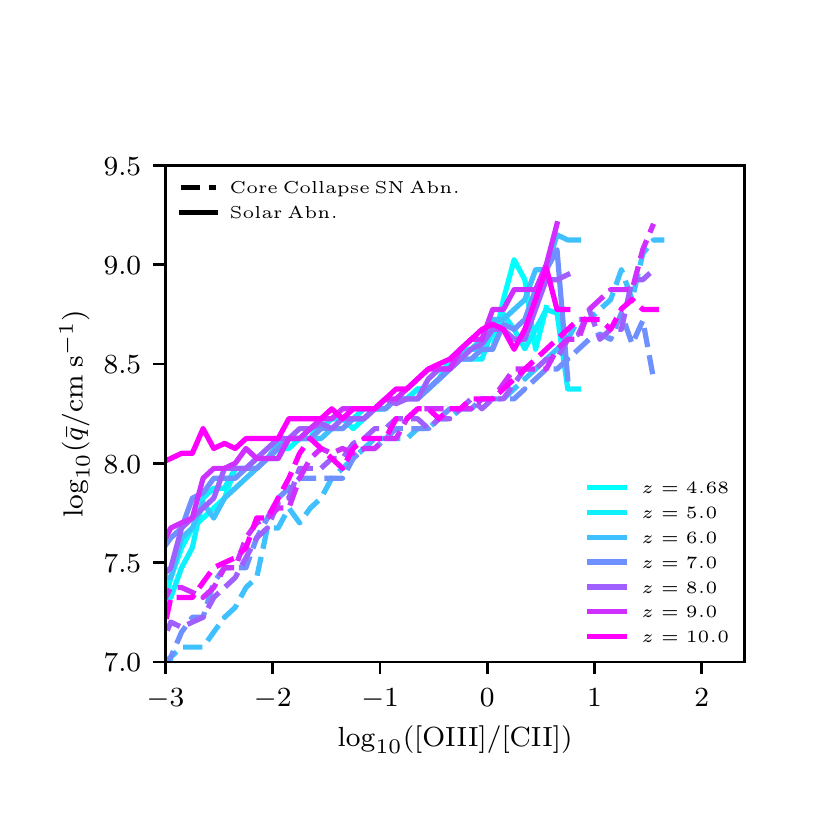}}
\caption{[OIII]/[CII] versus $\log_{10}(\bar{q})$ for galaxies in our simulation. $\bar{q}$ is defined to be the 90th percentile value from the HII regions in each galaxy. The coloured lines indicate the median relation for different redshift bins. Solid and dashed lines represent the results when we assume Solar abundances and core collapse SN abundances, respectively. We only include galaxies in this plot that have an ${\rm SFR}>10^{-2}{\rm M_{\odot}yr^{-1}}$.}
\label{CII-OIII-U}
\end{figure}

While it is trivial to set the ionisation parameter in a {\small CLOUDY} calculation, real galaxies exhibit huge ranges in $q$ depending on location within the galaxy. In order to describe the galaxy by a single value, we select all HII regions defined as gas with $T<10^{5.5}$K, $n_{\rm H}>1{\rm cm^{-3}}$ and $x_{\rm HII}>50\%$, and measure the 90th percentile mass-weighted ionisation parameter.  In the bottom right panel of Figure~\ref{2Dhists} we show a 2D histogram of [CII]/SFR versus [OIII]/SFR coloured by $\log_{10}(\bar{q})$. As predicted by \cite{Harikane2020}, we find the gradient in $q$ evolves towards the upper left of the figure (i.e. high ionisation parameter results in lower [CII]/SFR and higher [OIII]/SFR). This can be seen more clearly in Figure~\ref{CII-OIII-U} where we plot $\log_{10}(\bar{q})$ directly against [OIII]/[CII] for both the Solar abundance and core collapse SN abundance models. Ionisation parameters of galaxies with [OIII]/[CII] ratios of 10 can reach as high as $\log_{10}(\bar{q})\gtrsim9$. Therefore the higher [OIII]/[CII] ratios seen at high redshift could be partially due to higher ionisation parameters. 

\begin{figure}
\centerline{\includegraphics[scale=1,trim={0 0.4cm 0 1.5cm},clip]{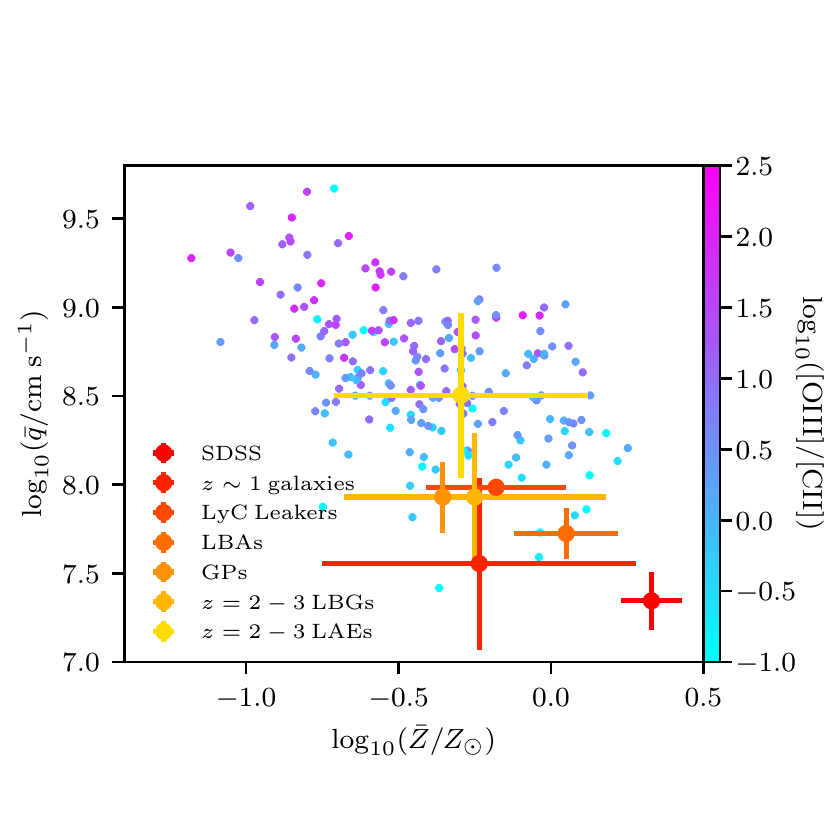}}
\caption{$\log_{10}(\bar{q})$ versus metallicity for galaxies in our simulation. $\bar{q}$ is defined to be the 90th percentile value from the HII regions in each galaxy whereas metallicity is the mass-weighted mean gas-phase metallicity in the ISM. We only show galaxies with ${\rm SFR > 1M_{\odot}/yr}$ for the core-collapse SN abundance model with enhanced metallicity. Data points with error bars represent ranges in ionisation parameter and metallicity for SDSS galaxies, $z\sim1$ galaxies, LyC leakers, Lyman-break analogues (LBAs), green pea (GP) galaxies, $z=2-3$ Lyman-break galaxies (LBGs), and $z=2-3$ Ly$\alpha$ emitters (LAEs) from \protect\cite{Nakajima2014}.}
\label{CII-OIII-U-Z}
\end{figure}

While measurements of the ionisation parameter at $z>6$ are not yet possible, proxies from lower-redshift may give an indication as to whether reionization epoch systems truly exhibit higher values of $q$. The O32 diagnostic (i.e. ${\rm [OIII]_{4960\si{\angstrom},5007\si{\angstrom}}/[OII]_{3727\si{\angstrom},3729\si{\angstrom}}}$) is often used as a proxy for ionisation parameter \citep[e.g.][]{Penston1990}. \cite{Strom2017} demonstrated using the KBSS survey that star-forming galaxies at $z\sim2-3$ exhibit considerably higher O32 compared to low-redshift SDSS galaxies, indicative of higher ionisation parameters. Similarly, \cite{Katz2019} showed that massive high-redshift galaxies in the epoch of reionization have very similar strong-line diagnostics compared to these lower-redshift systems. Consistent results were also found for LyC leakers in the redshift range $z\sim2-4$ \citep{Nakajima2020}. 

In Figure~\ref{CII-OIII-U-Z}, we compare {\small SPHINX$^{20}$} galaxies with ${\rm SFRs > 1M_{\odot}/yr}$ to different classes of observed galaxies on the ionisation parameter-metallicity plane assuming the core-collapse SN abundance model with enhanced metallicity. The vast majority of star-forming {\small SPHINX$^{20}$} galaxies exhibit ionisation parameters greater than that of SDSS galaxies, and often equal to or greater than the ionisation parameters in $z=2-3$ Ly$\alpha$ emitters or known LyC leakers. There is a clear trend that lower metallicity star-forming galaxies exhibit higher ionisation parameters. Furthermore, it is often the lower metallicity, higher ionisation parameter galaxies that exhibit the highest [OIII]/[CII] ratios. However, at fixed $q$, the [OIII]/[CII] seems to be relatively constant (this is explored further in the next section), hence we argue that $q$ is more important for setting the [OIII]/[CII] ratio compared to $Z$.

In summary, observational evidence combined with theoretical models suggests that high-redshift galaxies exhibit higher ionisation parameters which can contribute to the high ratios of [OIII]/[CII]. However, in the case of the solar abundance model, all of the observed high-redshift galaxies would need to exhibit ionisation parameters of order $\log(q)\sim9$ in order to reach the [OIII]/[CII] observed. The core collapse SN models can reach the observed ratios with a lower ionisation parameter.

\begin{figure}
\centerline{\includegraphics[scale=1,trim={0 0.4cm 0 1.5cm},clip]{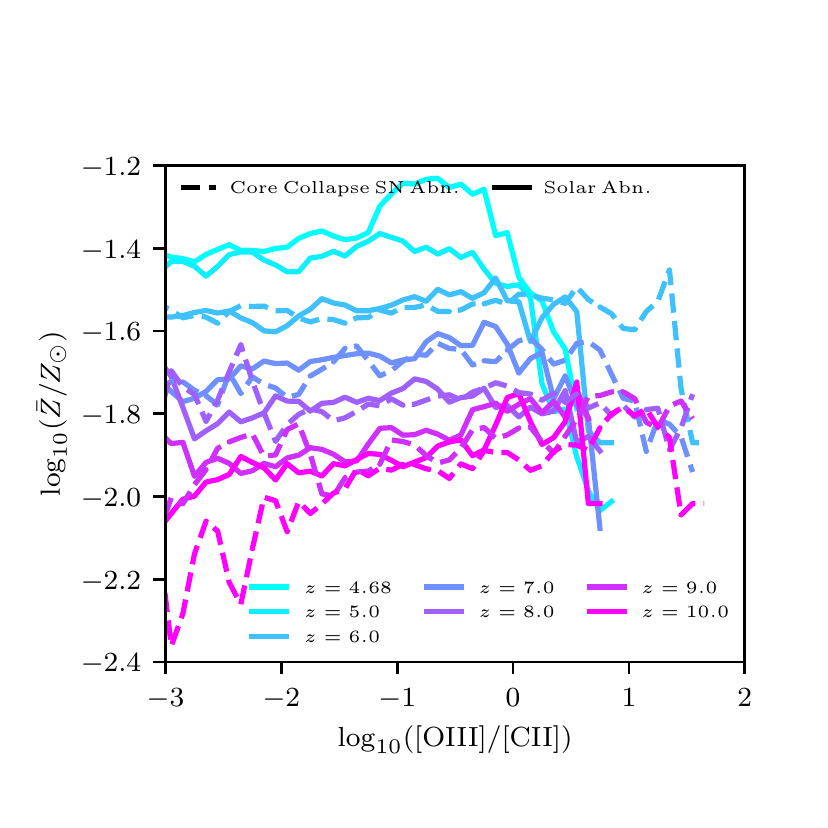}}
\caption{[OIII]/[CII] versus $\log_{10}(Z/Z_{\odot})$ for galaxies in our simulation. The coloured lines indicate the mean relation for different redshift bins. Solid and dashed lines represent the results when we assume Solar abundances and core collapse SN abundances, respectively. We only include galaxies in this plot that have an ${\rm SFR}>10^{-2}{\rm M_{\odot}yr^{-1}}$.}
\label{CII-OIII-Z}
\end{figure}

\subsubsection{Lower gas metallicities}
To first order, the line luminosity of each simulation cell scales with the amount of metals present in the gas. \cite{Harikane2020} argue that for lower metallicity gas, one expects a drop in [OIII]/SFR but a constant [CII]/SFR. Since the bulk of the [CII] emission arises from PDRs, the PDR column density is proportional to $1/Z$ if dust dominates the self-shielding from UV photons \citep{Kaufman2006}. For dwarf galaxies with ${\rm SFR}\geq1{\rm M_{\odot}yr^{-1}}$ in the local Universe, \cite{DeLooze2014} find an enhancement in [CII]/SFR for metal-poor dwarf galaxies compared to other galaxy types. 

Similar to ISM density, the [OIII]/[CII] ratio is sensitive to numerous different effects that either affect or are impacted by metallicity. For example, lower metallicity stars have harder SEDs which could result in the production of more O$^{++}$, while the additional ionising photons may ionise some of the PDRs that would be present otherwise. The abundance ratios are expected to change with metallicity and \cite{Arata2020} predicted from simulations that lower metallicity galaxies exhibit higher ratios of [OIII]/[CII].  The final amount of momentum injected during the SN snowplow phase is weakly sensitive to the metallicity of the gas \citep{Blondin1998,Thornton1998,Karbov2020} and similarly the ability for gas to cool is also sensitive to metallicity \citep[e.g.][]{Sutherland1993}. Finally, there is a well established mass-metallicity relation in the local Universe \citep{Tremonti2004} and total halo mass is partially responsible for setting the maximum densities achievable in our simulation.

In the bottom left panel of Figure~\ref{2Dhists}, we show a 2D histogram of [CII]/SFR versus [OIII]/SFR for galaxies in our simulation coloured by average gas-phase metallicity in the regions of the galaxies that emit [OIII] and [CII]. The higher metallicity galaxies evolve towards the top right of the diagram, consistent with the idea that to first order, [OIII] and [CII] luminosity scale with total metallicity. Note that our {\small CLOUDY} calculations do include dust, but we have chosen a dust-to-metal ratio that scales with metallicity \citep{RR14} so most of the galaxies are relatively dust free. In fact, our results are very similar when the {\small CLOUDY} models are run without dust. Therefore, we do not recover the trend predicted by \cite{Harikane2020}. Furthermore, as the abundance ratios in our simulation do not evolve with metallicity, we also do not recover the trend seen in \cite{Arata2020}, although we emphasise that their trend is likely more realistic and the evolution is most likely bounded by our two different abundance ratio models. Rather, metallicity in our simulation, in the context of the [OIII]/[CII] ratio, is a more sensitive indicator of the ISM physics and star formation history which helps set the emission line luminosities.


In Figure~\ref{CII-OIII-Z}, we plot the mean relations of [OIII]/[CII] versus $\log_{10}(Z/Z_{\odot})$ as a function of redshift. There is a strong evolution with decreasing redshift towards higher values of metallicity simply because as the simulation evolves, the galaxies become more metal enriched. The trend between [OIII]/[CII] and metallicity is essentially flat indicating that for dust-poor galaxies, excluding the impact of evolving abundance ratios with metallicity, in general, the metallicity of the galaxy does not strongly impact the [OIII]/[CII] ratio. We do however point out that for the galaxies with high SFRs, we indeed find a trend that lower metallicity galaxies tend to also exhibit higher ionisation parameters and higher [OIII]/[CII] ratios (see Figure~\ref{CII-OIII-U-Z}). However, at fixed $q$, the [OIII]/[CII] ratio seems constant. While in general, metallicity may not impact the relation for the majority of galaxies in our simulation, at higher SFRs, since there is a trend between ionisation parameter and metallicity, it may appear that low metallicity correlates with high [OIII]/[CII], opposite of what is suggested in \cite{Harikane2020}.

\subsubsection{Cosmic Microwave Background attenuation}
The CMB can have multiple effects on high-redshift infrared line emission because the radiation temperature is significantly higher at $z>6$ than it is in the local Universe. The CMB acts as both a background against which one has to observe the line and can also act as a heating term which is important for low temperature gas \cite[e.g.][]{Lagache2018}. While the CMB can impact [OIII], it is particularly important in the context of [CII], especially for lower density, low-temperature gas (see e.g. Figure~1 of \citealt{Kohandel2019}). Strong CMB attenuation of the [CII] line can drive higher ratios of [OIII]/[CII] and partially explain the elevated ratio at high-redshift compared to the local Universe.

Both \cite{Laporte2019} and \cite{Harikane2020} considered the impact of CMB attenuation and concluded that the effect was not strong enough to be completely responsible for the elevated [OIII]/[CII] ratios. \cite{Lagache2018} predicted that the maximum increase in [OIII]/[CII] due to CMB attenuation is $\sim0.5$~dex. This value is sensitive to both the temperature and density structure of the gas that emits [CII].  

To measure the impact of CMB attenuation, we have run an additional set of {\small CLOUDY} models\footnote{These {\small CLOUDY} models were run without dust and compared to a matched set of {\small CLOUDY} models with a $z=6$ CMB that also neglect dust.} for our simulated galaxies at $z=6$ where we have artificially set the CMB background to be consistent with the value at $z=0$, thus removing the CMB attenuation effects. Considering only galaxies that have [CII] and [OIII] luminosities of at least $10^3L_{\odot}$, for most of the systems at $z=6$, the enhancement in [OIII]/[CII] is only $\sim0.05$~dex and thus we concur with \cite{Laporte2019} and \cite{Harikane2020} that while CMB attenuation enhances the [OIII]/[CII] ratio slightly, it cannot be a dominant factor.

\subsubsection{Extended [CII] emission}
Using stacks of $z=5-7$ galaxies, \cite{Fujimoto2019} identified large haloes of [CII], extending out to $\sim10$kpc in galaxies with SFRs of $\sim10-70{\rm M_{\odot}yr^{-1}}$. Taking into account this extended emission, \cite{Carniani2020} determined that earlier observations may underestimate the [CII] luminosity by up to a factor of $\sim2$. Simulations have struggled in reproducing large [CII] haloes \citep[e.g.][]{Katz2019,Pallottini2019,Arata2020} such as those observed in \cite{Fujimoto2019}. Nevertheless, if the [CII] surface brightness profile is significantly more extended than that of [OIII], the underestimated values [CII] may lead to artificially high values of [OIII]/[CII].

\begin{figure}
\centerline{\includegraphics[scale=1,trim={0 0.4cm 0 1.5cm},clip]{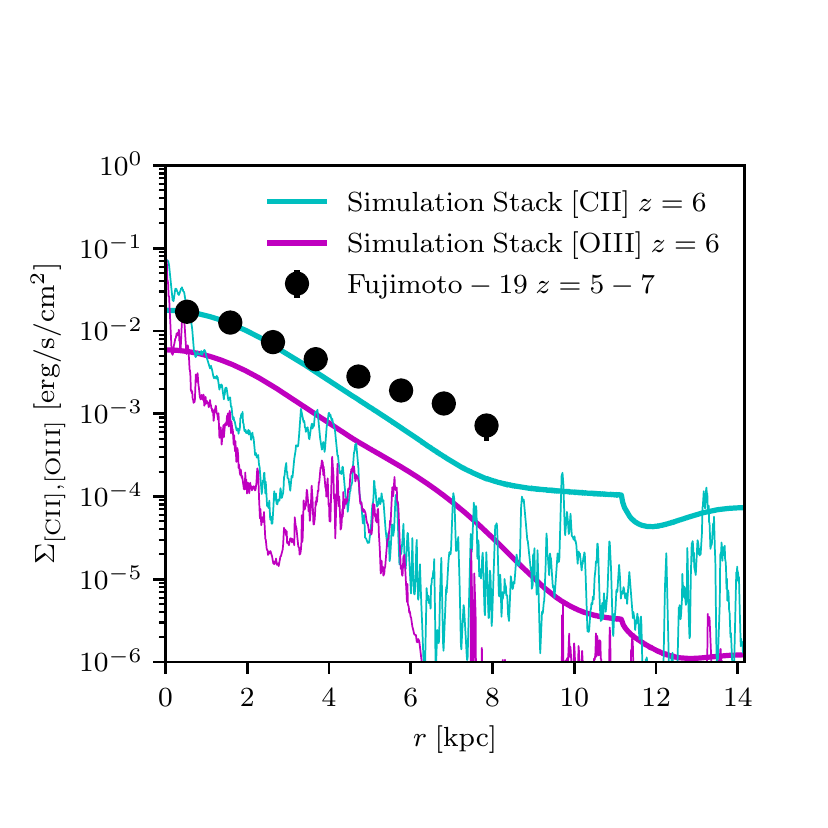}}
\caption{Stacked [CII] and [OIII] surface brightness profiles at $z=6$ for the three most massive galaxies in the simulation viewed along three different angles compared to an observed stack at $z=5-7$ from \protect\cite{Fujimoto2019}. The [OIII] profile is only slightly more concentrated than that of [CII]. The thick lines represent the profiles that have been smoothed with the ALMA beam while the thin lines represent the intrinsic stacked profile. All profiles have been normalised so that the [CII] profile matches the innermost data point from \protect\cite{Fujimoto2019}.}
\label{sb_prof}
\end{figure}

In Figure~\ref{sb_prof}, we plot the [OIII] and [CII] surface brightness profiles for a stack of the three most massive galaxies in our simulation at $z=6$ for three different viewing angles.  These are the galaxies shown in Figure~\ref{hero} and represent the most similar galaxies in our simulation to those observed in \cite{Fujimoto2019}. We show results for when the stacks are smoothed with the ALMA beam (thick lines) as well as the un-smoothed values (thin lines). The surface brightness of the stacks have been renormalised by a constant factor so that the [CII] surface brightness profiles match the innermost data point from \cite{Fujimoto2019}. While the [CII] surface brightness profile is slightly more extended than that of [OIII], consistent with observations from \cite{Carniani2020}, the difference is small. 

The smoothed profile is in good agreement with the data from \cite{Fujimoto2019} up to $\sim4.5kpc$ before dropping below the observational data points. Smoothing moves the intrinsic profile into significantly better agreement with observed data, but there is still a discrepancy at large radii. Part of the extended profile that we see is driven by mergers. In the second row of Figure~\ref{hero}, we can see that the second most massive halo at $z=6$ is a triple merger of three smaller galaxies. It is not particularly obvious how to choose a centre for stacking the images given the complex morphology and this can also play a role in the shape of the surface brightness profile. In Figure~\ref{sb_prof}, we have opted to use the centre of the halo.

Because the [OIII] surface brightness profile is so similar to that of [CII], it is unlikely that galaxy [OIII] and [CII] morphologies are responsible for the high [OIII]/[CII] ratios observed at high redshift, consistent with the arguments posed in \cite{Harikane2020} because even when accounting for surface brightness dimming in observations \citep{Carniani2020}, high-redshift galaxies still exhibit high [OIII]/[CII] ratios. 

\begin{figure}
\centerline{\includegraphics[scale=1,trim={0 0.5cm 0 0cm},clip]{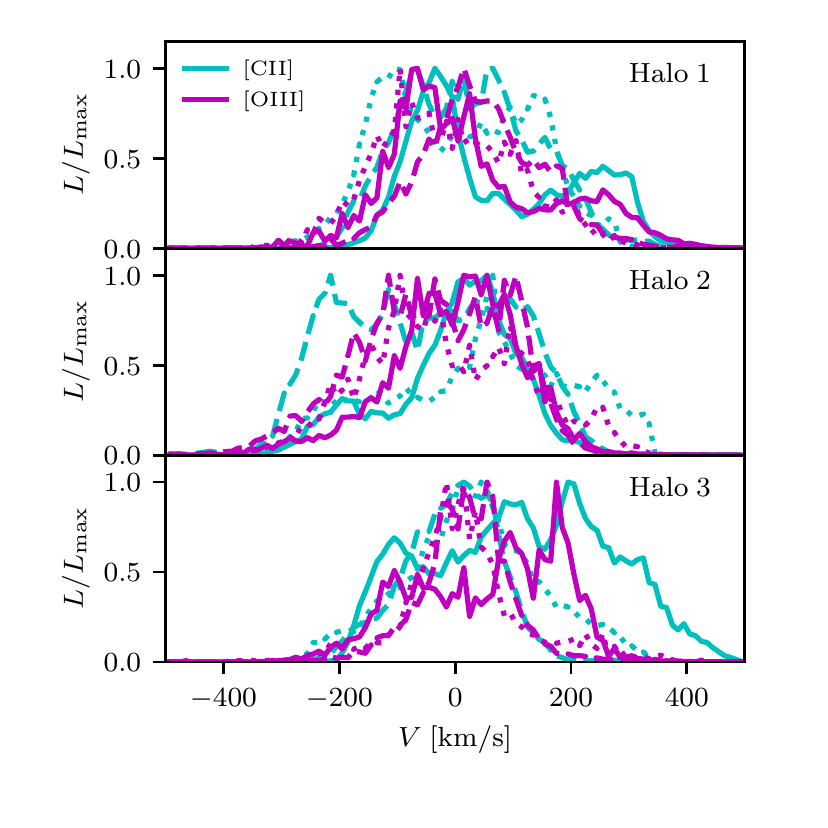}}
\caption{[CII] (cyan) and [OIII] (magenta) spectra for three different galaxies at $z=6$ for three different viewing angles (solid, dashed, and dotted lines). The shape of the spectra change depending on the viewing angle and it is not necessarily the case that the [OIII] and [CII] spectra exhibit the same shape. We show the spectra for the Solar abundance model (although the shape will be similar for the core collapse SN abundance model) and the spectra have been smoothed over bins of 20 km/s. We have normalised each spectrum to its maximum value so that the FWHM can be easily compared.}
\label{spectra}
\end{figure}

\subsubsection{Inclination effects}
\cite{Kohandel2019} demonstrated using hydrodynamics simulations that because the shape of the [CII] spectrum changes with viewing angle of a galaxy, for orientations that lead to large velocity spreads (e.g. edge-on) the [CII] flux may not be high enough to be detected by ALMA. In contrast, when the galaxy is viewed face-on, it is more likely to be detected. \cite{Harikane2020} speculate that this effect is unlikely to change the [OIII]/[CII] ratio because if [CII] is undetected due to orientation, [OIII] is also likely to be impacted in a similar way. However, this argument implicitly assumes that [CII] and [OIII] have similarly shaped spectra or that [OIII] is not always more centrally concentrated. \cite{Katz2019} showed that in certain cases, the first of these two assumptions is incorrect, especially during strong dynamical interactions or feedback events. This once again arises because [OIII] is emitted from a different gas phase than [CII]. 

In Figure~\ref{spectra} we show the [CII] and [OIII] spectra of the three most massive galaxies at $z=6$ for three different viewing angles. We have normalised each spectrum so the full width at half maximum (FWHM) can be more easily compared between the lines that have different luminosities. Consistent with \cite{Kohandel2019}, we also see strong variations in the spectra as a function of viewing angle as expected. While the [OIII] and [CII] spectra exhibit many similar features, the shapes are not identical. In some cases, the [CII] spectrum is more extended than that of [OIII] (see the solid lines in the top panel) which can lead to a [CII] deficit for this galaxy; however, this is not a generic feature.  For example, in the second row, the dotted line peak for [OIII] is nearly 200~km/s offset from the [CII] peak towards negative velocity. 

While galaxy orientation may explain [CII] deficits in specific high-redshift galaxies, it cannot explain the high [OIII]/[CII] ratios in all of the observed high-redshift systems, unless there is an observational bias that preferentially selects concentrated [OIII] emission or spatially extended [CII] emission.

\begin{figure*}
\centerline{\includegraphics[scale=1,trim={0 0cm 0 0cm},clip]{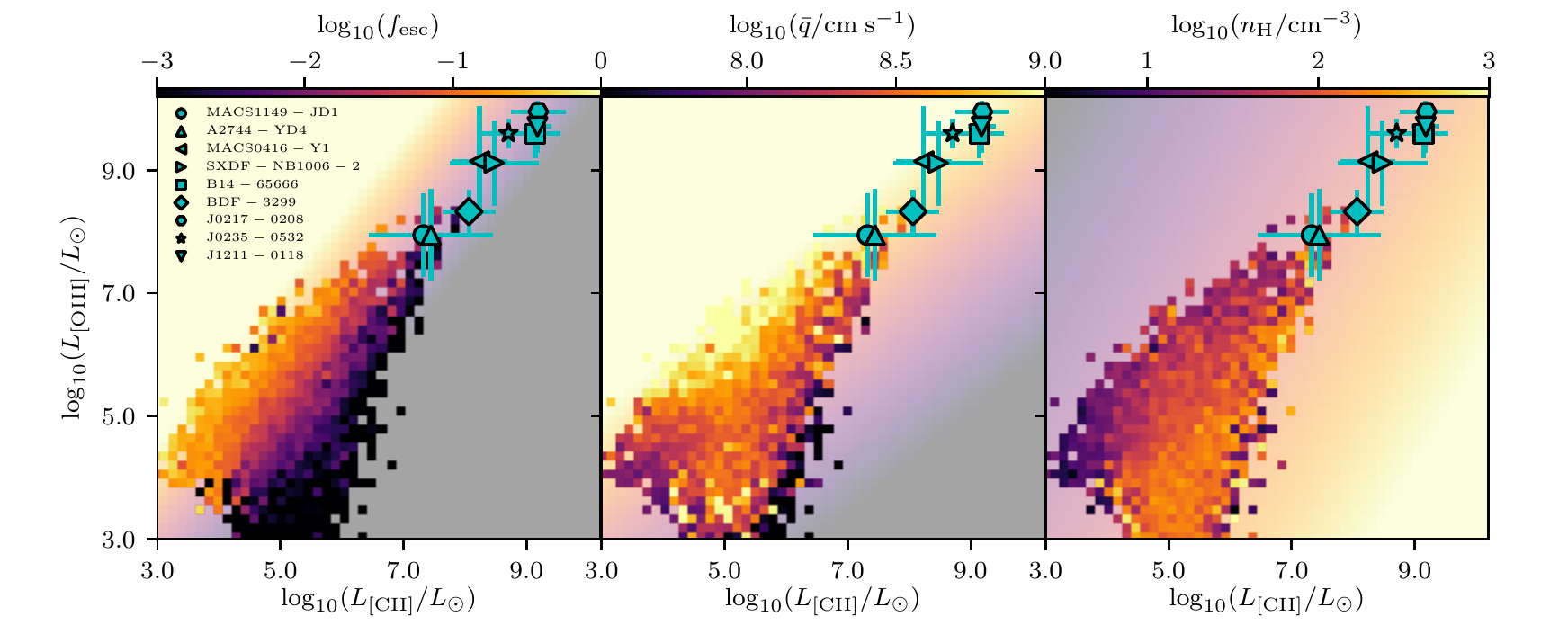}}
\caption{2D histograms of [CII] versus [OIII] coloured by the mean values of $f_{\rm esc}$ (left), 90th percentile ionisation parameter (centre), and 90th percentile ISM density (right) for galaxies in {\small SPHINX$^{20}$} at $6\leq z\leq10$. The underlaid colour map shows the results from the fitted generalised linear model for each parameter. We have shown the results for the model that assumes the core collapse SN abundances with the metallicity scaled by a factor of 4 as this best reproduces the high-redshift [CII]-SFR and [OIII]-SFR relations. For comparison, we show nine $z>6$ galaxies from \protect\cite{Carniani2020}.}
\label{gal_props}
\end{figure*}

\subsection{Alternative Explanations for High [OIII]/[CII] Ratios}
We have discussed eight different effects that can potentially drive high-redshift galaxies towards high [OIII]/[CII] ratios; however, there is still physics that is not considered in our model that may play a role in setting the [OIII]/[CII] ratios of high redshift galaxies. 

At slightly lower redshifts than considered here, it has been shown that the [CII] line is optically thick in some of the brightest submillimeter galaxies \citep[e.g.][]{Neri2014}. We can obtain a rough estimate of [CII] by measuring the optical depth to [CII] in every simulation cell.  Following \cite{Goldsmith2012,Lagache2018}, the optical depth ($\tau$) can be written as 
\begin{equation}
\tau=\tau_0\frac{1-e^{-T_*/T_{\rm ex}}}{1+\left(\frac{g_u}{g_l}\right)e^{-T_*/T_{\rm ex}}},
\end{equation}
where $T_*=91.25$K is the equivalent temperature of the transition, $g_u=4$ and $g_l=2$ are the statistical weights of the upper and lower levels, $T_{\rm ex}$ is the excitation temperature of the gas, and 
\begin{equation}
    \tau_0=7.49\times10^{-18}N_{\rm C^{+}}/\delta v,
\end{equation}
where $N_{\rm C^{+}}$ is the CII column density and $\delta v$ is the line width. The line width is estimated based on the temperature of the cell and the microturbulence of the gas such that $\delta v=\sqrt{a_{\rm therm}^2+a_{\rm turb}^2}$. Here, $a_{\rm therm}=\sqrt{\frac{2k_BT}{\mu m_p}}$ where $T$ is the temperature of the cell, $k_B$ is the Boltzmann constant, $m_p$ is the proton mass and $\mu$ is the mean weight of $^{12}{\rm C^+}$. To estimate the microturbulence, we follow \cite{Larson1981,Franeck2018} and set $a_{\rm turb}=1.1\times(l/{\rm pc})^{0.38}$, where $l$ is the physical length of a cell in the simulation. Based on our {\small CLOUDY} models, the fraction of C that is in the form of ${\rm C^+}$ is well described as a function of temperature such that
\begin{equation}
    f_{\rm C^+} = \frac{1}{1+e^{-(147.27-35.69\log_{10}(T/K))}}.
\end{equation}
$N_{\rm C^{+}}$ can then be calculated for every cell knowing the density, metallicity, $f_{\rm C^+}$ and cell size. We can measure the excitation temperature in each cell using the collisional deexcitation rate coefficients for neutral hydrogen and electrons as given in \cite{Goldsmith2012} and assuming a CMB background at $z=6$. To calculate the true extinction, one must then ray trace through the simulation accounting also for the bulk velocities of every gas cell. Rather, to obtain a simple estimate for the effect of self-absorption, we assume that the emission can only be absorbed locally and that nearby cells exhibit similar $\tau$.  Under this simple approximation, for the most massive galaxy at $z=6$, we find that $L_{\rm [CII]}$ is reduced from $5.24\times10^{41}{\rm ergs/s}$ to $2.62\times10^{41}{\rm ergs/s}$, which is equal to a 50\% decrease in total luminosity. We have completed this exercise for the next two most massive galaxies at $z=6$ and find similar reduction factors. Hence self-absorption can indeed be important for the most massive objects in the epoch of reionization; however, it is unlikely to be fully responsible for the high [OIII]/[CII] ratios seen, especially in the lower mass objects that are less metal enriched and when turbulence is high compared to the thermal velocity.

Another physical effect not considered here is different rates of depletion of C and O onto dust. While C and O are not expected to be heavily depleted like some other elements (e.g. Fe, \citealt{Jenkins2009}), if for example C is more depleted than O, we would expect that the [OIII]/[CII] ratio would be enhanced. Because our simulated galaxies have low metallicity, we expect the effect of depletion of metals onto dust to be small. 

We have hinted earlier that we also cannot rule out observational bias and the preferential selection of galaxies at high-redshift that either exhibit [CII] deficits or enhanced [OIII] emission. While observational bias does not explain why some high-redshift galaxies exhibit exceptionally high [OIII]/[CII] ratios, it may help explain why all of the nine observed high-redshift galaxies with both [OIII] and [CII] detections fall high compared to the local relation. Galaxies are often selected by means of UV dropout. Since UV photons are primarily emitted by young stars, at fixed halo mass, a galaxy is more likely than not to be observed during a burst of star formation. Since both ionising photons from young stars and SN feedback enhance [OIII] emission, we might expect that UV selection biases galaxies towards having high [OIII]/[CII]. \cite{Vallini2021}, using the model of \cite{Ferrara2019} also found that extreme gas conditions due to starbursts in UV bright galaxies lead to high [OIII]/[CII]. Furthermore, galaxies that are more disrupted by feedback are likely to have less dust attenuation which favours brighter magnitudes at fixed SFR. This in turn suggests that galaxies not selected in the UV may not exhibit [CII] deficits. While the sample is still small, we point out that two galaxies at $z>6$ that were selected based on blue, rest-frame optical colours for ALMA follow-up are more consistent with, or perhaps even above the local [CII]-SFR relation \citep{Smit2018}. We do note that these galaxies may host AGN which have not been explicitly modelled in our work.

Finally, our models only consider stars as the sources of ionising radiation; however, if there are a substantial number of accreting black holes, either low-mass or obscured, they may evade observational detection while providing additional photons and feedback needed to increase [OIII].  Follow-up observations of additional lines, such as [NV] may help constrain this scenario.

\section{Discussion}

\subsection{Predicting Galaxy Properties with [OIII] and [CII]}
As was shown in Section~\ref{results}, the [OIII]/[CII] ratio is sensitive to many galaxy properties including the LyC escape fraction, ionisation parameter, ISM density, SFR, etc. [CII] and [OIII] are already used as SFR indicators \citep[e.g.][]{DeLooze2014} whereas \cite{Katz2020} showed that the combination of [OIII] and [CII] could be used to identify LyC leakers. Here, we aim to constrain the individual properties of high-redshift galaxies using the correlations found in {\small SPHINX$^{20}$}. 

In Figure~\ref{gal_props}, we show 2D histograms of [CII] versus [OIII] where we have coloured each individual pixel by the mean value of $f_{\rm esc}$ (left), 90th percentile HII region ionisation parameter (centre), and 90th percentile HII region density (right) for simulated galaxies at $6\leq z \leq 10$. The trends that were discussed in Section~\ref{results} are immediately visible. For example, the bias towards higher $f_{\rm esc}$ and $q$ at high [OIII] for fixed [CII]. We show the model for the core collapse SN abundances with the 4 times enhanced metallicity as this is the model that best reproduces the [CII]-SFR and [OIII]-SFR relations at high redshift. We have fit relations between each parameter and [CII] and [OIII] using a generalised linear model such that:
\begin{equation}
\begin{split}
    \log_{10}(f_{\rm esc}) = &\ {\rm Min[Max[ }-0.91 + 0.95\log_{10}\left(\frac{L_{\rm [OIII]}}{L_{\odot}}\right) \\ &\ -1.11\log_{10}\left(\frac{L_{\rm [CII]}}{L_{\odot}}\right),-10],0],
\end{split}
\end{equation}
\begin{equation}
\begin{split}
    \log_{10}\left(\frac{q}{\rm cm\ s^{-1}}\right) = &\ {\rm Min[Max[}8.66 + 0.27\log_{10}\left(\frac{L_{\rm [OIII]}}{L_{\odot}}\right) \\ 
    &\ -0.29\log_{10}\left(\frac{L_{\rm [CII]}}{L_{\odot}}\right),7.5],9.5],
\end{split}
\end{equation}
and
\begin{equation}
\begin{split}
    \log_{10}\left(\frac{n_{\rm H}}{{\rm cm^{-3}}}\right) = & \ {\rm Min[Max[}1.63-0.18\log_{10}\left(\frac{L_{\rm [OIII]}}{L_{\odot}}\right) \\ &\  +0.21\log_{10}\left(\frac{L_{\rm [CII]}}{L_{\odot}}\right),0],4].
\end{split}
\end{equation}
We have defined these relations over fixed intervals for each parameter by the Min and Max functions in each equation. We have underlaid each of these relations in Figure~\ref{gal_props} where one can see that the generalised linear models provide an adequate representation of the simulated galaxies. To determine the uncertainty of these relations, we calculate the expected $f_{\rm esc}$, $q$ and $n_{\rm H}$ for each of the simulated galaxies and measure the difference between the true and predicted values. The 50th percentile errors on the parameters are 0.006 for $f_{\rm esc}$, 0.47~dex for $n_{\rm H}$, and 0.29~dex for $q$.

Also shown in Figure~\ref{gal_props} are nine $z>6$ galaxies with [CII] and [OIII] detections from \cite{Carniani2020}. Six of the nine galaxies have [CII] and [OIII] luminosities that fall either outside or very close to the end of the regions of the [CII]-[OIII] plane probed by {\small SPHINX$^{20}$}, while three galaxies, including MACS1149-JD1, the highest redshift galaxy in the sample, fall well within range of the simulation. 

\begin{table*}
    \centering
        \caption{Estimated ISM and HII region properties for nine $z>6$ galaxies. Redshifts, [CII], and [OIII] luminosities are the surface brightness dimming corrected values from \protect\cite{Carniani2020} while $f_{\rm esc}$, ionisation parameter and density and predicted from our models. Ionisation parameter and density are estimated for HII regions.}
    \begin{tabular}{lcccccc}
     \hline
     Galaxy & Redshift & $L_{\rm [CII]}/10^8L_{\odot}$ & $L_{\rm [OIII]}/10^8L_{\odot}$ &  $f_{\rm esc}$ & $\log_{10}(\bar{q}/{\rm cm\ s^{-1}})$ & $\log_{10}\left(\frac{n_{\rm H}}{{\rm cm^{-3}}}\right)$ \\
\hline
${\rm MACS1149-JD1}$ & 9.11 & $0.21\pm0.05$ & $0.88\pm0.19$ & $0.021\pm0.009$ & $8.79\pm0.04$ & $1.9\pm0.04$ \\ 
${\rm A2744-YD4}$ & 8.38 & $0.28\pm0.09$ & $0.88\pm0.21$ & $0.017\pm0.021$ & $8.77\pm0.05$ & $1.93\pm0.05$ \\ 
${\rm MACS0416-Y1}$ & 8.31 & $1.73\pm0.25$ & $14.0\pm4.0$ & $0.022\pm0.007$ & $8.92\pm0.05$ & $1.96\pm0.03$ \\ 
${\rm SXDF-NB1006-2}$ & 7.21 & $3.0\pm0.7$ & $13.1\pm2.9$ & $0.012\pm0.005$ & $8.86\pm0.04$ & $2.03\pm0.04$ \\ 
${\rm B14-65666}$ & 7.16 & $13.9\pm1.8$ & $39.0\pm5.0$ & $0.005\pm0.001$ & $8.84\pm0.02$ & $2.14\pm0.02$ \\ 
${\rm BDF-3299}$ & 7.11 & $1.16\pm0.16$ & $2.13\pm0.24$ & $0.006\pm0.001$ & $8.73\pm0.02$ & $2.04\pm0.02$ \\ 
${\rm J0217-0208}$ & 6.2 & $15.6\pm2.2$ & $89.3\pm2.1$ & $0.01\pm0.002$ & $8.94\pm0.02$ & $2.1\pm0.02$ \\ 
${\rm J0235-0532}$ & 6.09 & $5.1\pm0.8$ & $39.9\pm3.1$ & $0.017\pm0.003$ & $8.95\pm0.02$ & $2.02\pm0.02$ \\ 
${\rm J1211-0118}$ & 6.03 & $15.1\pm1.1$ & $51.0\pm7.0$ & $0.006\pm0.001$ & $8.87\pm0.02$ & $2.13\pm0.01$ \\ 
    \hline
    \end{tabular}
    \label{props_tab}
\end{table*} 

In Table~\ref{props_tab}, we list the values of $f_{\rm esc}$, $q$, and $n_{\rm H}$ for each of the nine observed high-redshift galaxies as predicted by our model. Interestingly, these galaxies all have $f_{\rm esc}\lesssim5\%$, which is below the luminosity weighted values at all redshifts in {\small SPHINX$^{20}$} and more consistent with the globally averaged value of $f_{\rm esc}$ (Rosdahl~et~al~{\it in prep.}). Such values are typical of more massive galaxies at $z>6$ in our simulations. These nine galaxies all have 90th percentile ionisation parameters with $\log_{10}(\bar{q}/{\rm cm\ s^{-1}})>8.5$ and 90th percentile HII region densities of $\log_{10}(n_{\rm H}/{\rm cm^{-3}})>1.5$. The LyC escape fractions scatter below that expected from the relation in Figure~\ref{CII-OIII-desc} due to the mass dependence of this quantity. The relations presented in Section~\ref{results} are weighted by number of galaxies rather than luminosity and are thus dominated by lower mass systems compared to what has been observed.

Compared to lower redshift galaxies, we find that our estimates for the ionisation parameters of the observed high-redshift galaxies tend to be more typical of $z\sim2-3$ Ly$\alpha$ emitting galaxies rather than local galaxies from SDSS (see Table~3 of \citealt{Sanders2015} and Figure~\ref{CII-OIII-U-Z}). We emphasise that our values of $q$ and $n_{\rm H}$ may not be directly comparable to those constrained by observations because of the different methods of measuring ionisation parameter. Our values represent the mass-weighted values of $q$ and $n_{\rm H}$ in HII regions which is represented by gas with $n_{\rm H}>1{\rm cm^{-3}}$, $T<10^{5.5}$K and $x_{\rm HII}>50\%$. 

We highlight that estimates of gas density in these high-redshift galaxies are not always consistent between different methods in the literature. For example for MACS1149\_JD1, \cite{Yang2020} find an upper limit gas density of $\log_{10}\left(n_{\rm H}/{\rm cm^{-3}}\right)=2.4-2.8$ or $1.6-2.0$ depending on how scatter between HII regions is accounted for, whereas \cite{Vallini2021} estimate a value of $0.88^{+ 0.55}_{-0.19}$. We find a 90th percentile HII region (log) density of 1.69 for the same galaxy. These density estimates are not all measuring the exact same quantity. The \cite{Vallini2021} values represent the average over the [OIII] and [CII] emitting regions while the \cite{Yang2020} values represent the upper limits in HII regions. Our 90th percentile values are perhaps more comparable to the upper limits quoted in \cite{Yang2020}. Consistent with \cite{Vallini2021}, we find that including both HII regions and PDRs in the calculation increases the density estimate.

\subsection{The Case for [OI] Observations}
Much of the complexity in interpreting ISM properties in real high-redshift galaxies using [OIII] and [CII] is that without knowing the star formation history and metallicity of the system, the different timescales for O and C enrichment can impact the determination of galaxy properties. Values of $f_{\rm esc}$, $q$, and $\rho$ are all sensitive to the C/O abundance ratio assumed. One potential way to circumvent this issue is to focus on line ratios that are not as impacted by the exact metal abundance ratios.   

The [OI] 63$\mu$m line has many similar properties to [CII] in that it is predominantly emitted by neutral gas since the ionisation energy of OI is 13.6eV (nearly exactly that of HI). [OI] has a very high critical density ($\sim5\times10^5{\rm cm^{-3}}$) and is known to strongly correlate with star formation in the local Universe \citep{DeLooze2014}. Numerical simulations predict that [OI] can outshine [CII] at high redshift \citep[e.g.][]{Olsen2017,Katz2019} and an [OI]/[CII] ratio $>1$ was recently detected with ALMA at $z>6$ \citep{Rybak2020}. [OI] is however impacted by self-absorption \citep{Kraemer1996,Goldsmith2019}. The primary benefit of observing [OI] is that the comparison with [OIII] should not be as impacted by different enrichment channels.

\begin{figure}
\centerline{\includegraphics[scale=1,trim={0 0cm 0 0cm},clip]{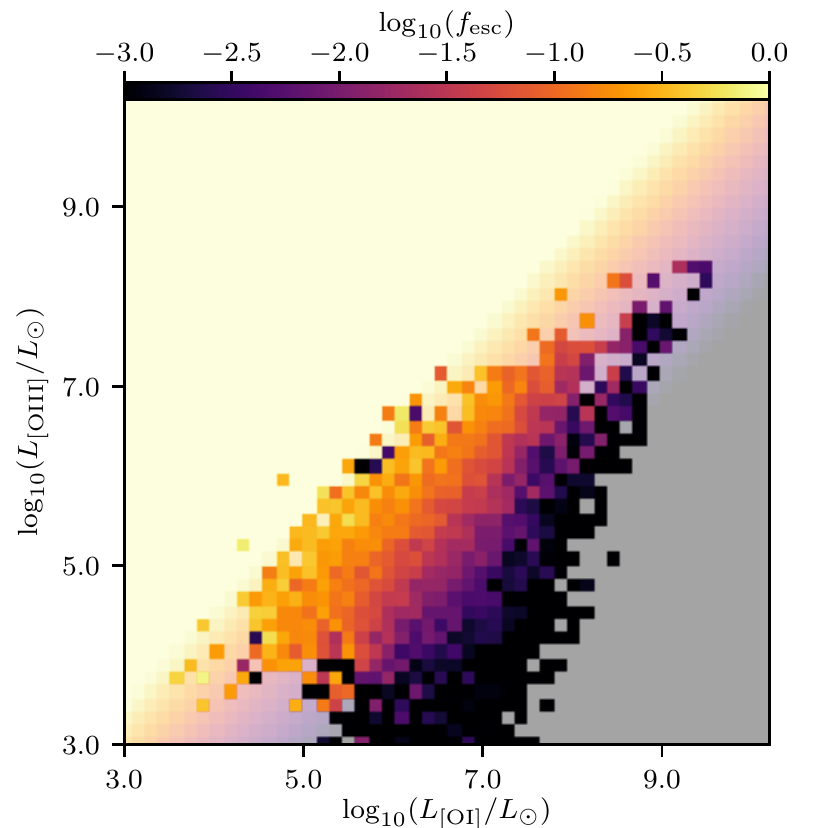}}
\caption{2D histogram of [OI] versus [OIII] coloured by the mean value of $f_{\rm esc}$ for galaxies in {\small SPHINX$^{20}$} at $6\leq z\leq10$. The underlaid colour map shows results from the fitted generalised linear model. We have shown the results for the model that assumes core collapse SN abundances with the metallicity scaled by a factor of 4.0.}
\label{OIII_OI_fesc}
\end{figure}

In Figure~\ref{OIII_OI_fesc} we show a 2D histogram of [OI] versus [OIII] coloured by the mean value of $f_{\rm esc}$ for galaxies in {\small SPHINX$^{20}$} at $6\leq z\leq10$. Similar to [CII], galaxies with higher [OIII] at fixed [OI] exhibit higher escape fractions due to the fact that [OI] originates in predominantly neutral gas whereas [OIII] is emitted by ionised gas. The relation between [OIII], [OI], and $f_{\rm esc}$ is well defined by the generalised linear model:
\begin{equation}
\begin{split}
    \log_{10}(f_{\rm esc}) = &\ {\rm Min[Max[ }-0.31 + 1.02\log_{10}\left(\frac{L_{\rm [OIII]}}{L_{\odot}}\right) \\ &\ -1.07\log_{10}\left(\frac{L_{\rm [OI]}}{L_{\odot}}\right),-10],0].
\end{split}
\end{equation}

The coefficients in this equation are nearly identical to the relation between [OIII], [CII], and $f_{\rm esc}$, while the intercept has increased due to the fact that [OI] is brighter. Given the utility of [OI] as a supplement or replacement for [CII], we argue that more observation time should be used to constrain [OI] emission from $z>6$ galaxies.

\subsection{Caveats}
\label{caveats}
As with all numerical simulations, there are numerous caveats we must be aware of when interpreting our results. We have modelled the emission lines in post-processing with {\small CLOUDY} and our method for setting up the models can introduce systematic uncertainties (see e.g. \citealt{Olsen2018}). For example, we have assumed a gas slab being irradiated from one direction whereas one could assume a shell instead. Simulation cells that host star particles are more likely to act like shells whereas gas cells without stars are more similar to slabs. Furthermore, we employ a constant temperature assumption to account for shocks and SN feedback that are present in the simulation rather than allow {\small CLOUDY} to compute the photoionisation-equilibrium temperature of the gas. Such an assumption can lead to different line luminosities as \cite{Lupi2020} showed that a variable temperature {\small CLOUDY} model can increase the [OIII]/[CII] ratio by a factor of two. One could also choose a different shape for the input SED than used in this work which can impact both the ionisation states of individuals as well as the electron density. Rather than model a different SED for each gas cell, we have fixed the shape of the SED within each radiation bin to be consistent with the Milky Way ISM and have normalised the photon count in each photon energy bin to match the simulation to conserve energy. The choice of SED within the cosmological simulation itself is also uncertain and it is well established that depending on the choice, the number of ionising photons as well as the escape fraction can change significantly \citep[e.g.][]{Ma2016,Rosdahl2018}. The SED will also vary with IMF as well as metallicity. Our results may change if the model with a top-heavy IMF and enhanced metallicity is self-consistently followed in the simulation. During post-processing, we also modelled the sub-ionising bands in the optically thin limit when in reality, molecules like H$_2$ or dust particles can absorb such photons and perhaps change the C$^+$ content of the galaxy. Finally, additional physics such as cosmic rays, magnetic fields, and subgrid turbulence are not included in the calculation, all of which can impact the emergent luminosities.

Our simulations are limited by both spatial and mass resolution which prevents us from modelling the detailed physics of the ISM on scales that we do not resolve. While our resolution is state-of-the-art in the context of full-box radiation hydrodynamics simulations, we do not resolve the formation of individual stars and their individual HII regions/PDRs where [OIII] and [CII] are emitted. Rather, we represent stellar clusters and therefore the structure of the ISM in the simulation may be different than in high-redshift galaxies. Furthermore, because we use a quasi-Lagrangian scheme for refining the AMR grid, most of our resolution tends to be concentrated on regions of high density. The origin of large [CII] haloes at high redshift is currently unknown and one suggestion is that there are small clumps of gas or small satellite galaxies in the outer regions of the haloes that emit [CII] \citep[e.g.][]{Fujimoto2019}. Simulations that add additional resolution to the CGM \citep[e.g.][]{Hummels2019,vdv2019} tend to resolve significantly more of these clumps.

Finally, we have assumed specific subgrid models for star formation and stellar feedback, all of which impact the stellar mass, metal enrichment, and ionising photon production. As we have argued, a more top-heavy stellar IMF which leads to enhanced SN feedback, ionising photon production, and metal enrichment may go a long way to reconciling the high-redshift [CII]-SFR, [OIII]-SFR, and [OIII]/[CII]-SFR relations. However, abundance ratios change for different mass stars and modifications to the assumed metal abundances ratios will result from a change in the stellar IMF. Enhancements in metallicity will also lead to additional cooling which is not completely self-consistent with our post-processing method. Our goal with this work is not to demonstrate that an exact metallicity increase of a factor of 4 is needed. Rather, we simply mean to demonstrate that increasing the metallicity from what is in our fiducial SPHINX model will push the simulated galaxies in the direction of observations. Future simulations will be needed to study these effects.

\section{Conclusions}
\label{conclusion}
In this work, we have studied the possible origins of high [OIII]/[CII] ratios observed in $z>6$ galaxies. We have used a new, state-of-the-art, full-box, cosmological radiation hydrodynamics simulation -- {\small SPHINX$^{20}$} -- to model [CII] and [OIII] for more than 20,000 simulated star-forming galaxies at $z>6$, allowing us to disentangle the physics that drives high [OIII]/[CII] ratios. Our main conclusions can be summarised as follows:

\begin{itemize}
    \item We can only reproduce the observed [CII]-SFR, [OIII]-SFR, and [OIII]/[CII]-SFR relations at $z>6$ if we assume that high-redshift galaxies are primarily enriched by low-metallicity core collapse SN and that the high-redshift stellar IMF is significantly more top heavy than the Kroupa IMF. This is consistent with expectations for low-metallicity star formation in the early Universe.
    
    \item The [CII]-SFR relation naturally converges to the observed values at $z\sim4.5$ when Solar abundance ratios with a standard IMF are assumed, indicating that both the stellar IMF and enrichment channels evolve significantly as redshift decreases.
    
    \item While the reduced C/O ratios at $z>6$ may be the primary reason why high-redshift galaxies exhibit high [OIII]/[CII] ratios, increased ionisation parameters, increased ISM density, increased LyC escape fractions, and to a small extent CMB attenuation also contribute to the high [OIII]/[CII] ratios. In contrast, lower gas metallicities, extended [CII] emission, and inclination effects are unlikely to explain the behaviour for all of the nine high-redshift galaxies with observations of both [CII] and [OIII].
    
    \item ${\rm [OIII]}$ emitting regions have significantly lower ISM densities than regions that emit [CII] and thus 1D models that consider a fixed gas density must be treated with caution when being used to study the ISM of the early Universe.
    
    \item ${\rm [OIII]}$ and [CII] can be used to constrain  ISM properties and LyC escape fractions of high-redshift galaxies. We find that typical escape fractions of the observed systems are $\sim1\%-4\%$ while gas densities in the [OIII] and [CII] emitting regions can reach as high as $>1000{\rm cm^{-3}}$. 
    
    \item ${\rm [OI]}$ can be used as both an alternative and supplement to [CII] observations in the high-redshift Universe because not only is it predicted to be brighter than [CII], it also originates from similar regions of the ISM meaning that it can be used to constrain similar galaxy characteristics as [CII]. We argue that more observation time should be spent on [OI] observations of $z>6$ galaxies.
    
\end{itemize}

Our model for predicting emission lines from simulations is applicable to numerous other lines across the wavelength spectrum. In future work we aim to compare how the methods for predicting galaxy properties such as metallicity, ionisation parameter, and gas density from strong-line diagnostics translate to the early Universe.

\section*{Acknowledgements}
We thank the anonymous referee for their comments which improved the manuscript. This work was supported by the Programme National Cosmology et Galaxies (PNCG) of CNRS/INSU with INP and IN2P3, co-funded by CEA and CNES. MGH  acknowledges support from the UKRI  Science and Technology Facilities Council (grant numbers ST/N000927/1 and ST/S000623/1). Support by ERC Advanced Grant 320596 “The Emergence of Structure during the Epoch of reionization” is gratefully acknowledged. SMA acknowledges support by ERC Starting Grant 638707 “Black holes and their host galaxies: co-evolution across cosmic time”. TK was supported by the National Research Foundation of Korea (NRF-2019K2A9A1A0609137711 and NRF-2020R1C1C100707911). TG acknowledges support from the European Research Council under grant agreement ERC-stg757258 (TRIPLE). The research of AS and JD is supported by Adrian Beecroft and STFC. Computing time for this work was provided by the Partnership for Advanced Computing in Europe (PRACE) as part of the ``First luminous objects and reionization with SPHINX (cont.)'' (2016153539, 2018184362, 2019215124) project. We thank Philipp Otte and Filipe Guimaraes for helpful support throughout the project and for the extra storage they provided us. We also thank GENCI for providing additional computing resources under GENCI grant A0070410560. RSE was supported by the European Research Council under the European Union Horizon 2020 research and innovation programme (grant agreement no. 669253).

\section*{Data Availability}
The data underlying this article will be shared on reasonable request to the corresponding author.

\bibliographystyle{mnras}
\bibliography{example}




\bsp	
\label{lastpage}
\end{document}